\documentclass[%
reprint, 
amsmath,amssymb,
aps,
prx,
  floatfix,
  longbibliography,
]{revtex4-2}

\usepackage{graphicx}
\usepackage{dcolumn}
\usepackage{bm}
\usepackage{physics}
\usepackage[usenames, dvipsnames]{color}
\usepackage{sidecap}
\usepackage{textcomp}
\usepackage{textgreek}
\usepackage{amsmath}
\usepackage{xr}

\definecolor{NewBlue}{rgb}{0, 0, 0.41}
\definecolor{NewRed}{rgb}{0.6, 0.07, 0.07}
\usepackage[colorlinks,
    linkcolor=NewBlue,
    citecolor=NewBlue,
    urlcolor=NewRed]{hyperref}

\begin{document}

\title{Universal high-fidelity quantum gates for spin-qubits in diamond}

\author{H. P. Bartling$^{1,2}$}
\thanks{These authors contributed equally to this work.}
\author{J. Yun$^{1,2}$}
\thanks{These authors contributed equally to this work.}
\author{K. N. Schymik$^{1,2}$}
\author{M. van Riggelen$^{1,2}$}
\author{L. A. Enthoven$^{1,3}$}
\author{H. B. van Ommen$^{1,2}$}
\author{M. Babaie$^{1,3,4}$}
\author{F. Sebastiano$^{1,3}$}
\author{M. Markham$^5$} 
\author{D. J. Twitchen$^5$}
\author{T. H. Taminiau$^{1,2}$}
\email{T.H.Taminiau@TUDelft.nl}

\affiliation{$^{1}$QuTech, Delft University of Technology, PO Box 5046, 2600 GA Delft, The Netherlands}
\affiliation{$^{2}$Kavli Institute of Nanoscience Delft, Delft University of Technology, PO Box 5046, 2600 GA Delft, The Netherlands}
\affiliation{$^{3}$Department of Quantum and Computer Engineering, Delft University of Technology, 2628 CJ Delft, The Netherlands}
\affiliation{$^{4}$Department of Microelectronics, Delft University of Technology, 2628 CD Delft, The Netherlands}
\affiliation{$^{5}$Element Six, Fermi Avenue, Harwell Oxford, Didcot, Oxfordshire, OX11 0QR, United Kingdom}

\date{\today}

\begin{abstract}

Spins associated to solid-state colour centers are a promising platform for investigating quantum computation and quantum networks. Recent experiments have demonstrated multi-qubit quantum processors, optical interconnects, and basic quantum error correction protocols. One of the key open challenges towards larger-scale systems is to realize high-fidelity universal quantum gates. In this work, we design and demonstrate a complete high-fidelity gate set for the two-qubit system formed by the electron and nuclear spin of a nitrogen-vacancy center in diamond. We use gate set tomography (GST) to systematically optimise the gates and demonstrate single-qubit gate fidelities of up to $99.999(1)\%$ and a two-qubit gate fidelity of $99.93(5) \%$. Our gates are designed to  decouple unwanted interactions and can be extended to other electron-nuclear spin systems. The high fidelities demonstrated provide new opportunities towards larger-scale quantum processing with colour-center qubits.

\end{abstract}

\maketitle

Solid-state colour centers are a promising qubit platform for exploring quantum simulation, computation and networks \cite{Awschalom2018}. Pioneering experiments include the demonstration of small quantum networks \cite{Hensen2015,Pompili2021,Hermans2022,Knaut2023}, control of quantum processors with up to 10 qubits \cite{Bradley2019,Nguyen2019,Bourassa2020,Randall2021}, the implementation of rudimentary quantum algorithms \cite{Vorobyov2021,vanDam2019,VanderSar2012,Wang2015,Zhang2020}, and the benchmarking of fault-tolerant quantum error correction codes \cite{Cramer2016,Waldherr2014,Abobeih2022}. 

One of the main challenges on the road to larger-scale quantum information processing is to further improve the fidelity of quantum gates. In the short term, the gate fidelity sets the attainable depth and complexity of quantum algorithms. In the longer term, error correction and fault tolerance can be used to overcome imperfections, provided that single- and two-qubit gate fidelities satisfy error thresholds \cite{Knill2005,Aliferis2006,Terhal2015,debone2024}. These prospects have stimulated a wide variety of research aiming to design, characterise and optimise high-fidelity quantum gates in various platforms \cite{VanderSar2012,Rong2015,Bourassa2020,Vallabhapurapu2023,Xie2023, Huang2019,Mills2022,Noiri2022,Xue2022,Madzik2022,Moskalenko2022,Kandala2021,Sung2021,Madjarov2020,Evered2023,Scholl2023,Wang2020,Ballance2016,Clark2021,Shukla2023}.  

For colour center qubits, the state of the art is set by the nitrogen-vacancy (NV) center in diamond, which can be used as a two-qubit system consisting of the NV electron spin and the intrinsic $^{14}$N nuclear spin \cite{VanderSar2012,Rong2015,Bourassa2020,Vallabhapurapu2023,Xie2023,Bradley2019}. High-fidelity single-qubit gates have been demonstrated for the NV electron spin and were characterised by randomized benchmarking \cite{Rong2015,Vallabhapurapu2023}. Recent work reported a high two-qubit electron-nuclear gate fidelity ($99.92 \%$) using gate designs that are tailored to the noise environment \cite{Xie2023}. However, the employed method of subspace randomized benchmarking does not consider the full two-qubit system and provides an upper-limit rather than a best-estimate for the fidelity \cite{Xie2023}. Additionally, that work did not include single-qubit gates on the nuclear-spin qubit, which are challenging to perform due to the always present electron-nuclear coupling.

Here, we design, optimise and characterise a complete set of high-fidelity quantum gates on the two-qubit system formed by the NV electron and nitrogen spins. We design gates that precisely decouple unwanted interactions between the two qubits, as well as between the qubits and the environment, and use gate set tomography (GST) \cite{Nielsen2021,Blume-Kohout2022} to obtain the process matrix for both single- and two-qubit gates in the full two-qubit space. We then demonstrate high fidelities for all gates in the two-qubit space (single-qubit gates around $F = 99.94 \%$ and two-qubit gate $F = 99.93(5) \%$). Additionally, we design and characterise single-qubit gates for specific cases in which one of the qubits is not actively used, and demonstrate a fidelity of $99.991(4) \%$ for the electron-spin qubit and a fidelity of $99.999(1)\%$ for the nitrogen-spin qubit. Finally, we compile a SWAP gate from the characterised gate set to illustrate how the information of the gate errors can be used to implement error mitigation techniques. We use this SWAP gate to store quantum states in the nitrogen quantum memory for over 100 seconds. Our systematic optimisation and high gate fidelities for a complete gate set provide a promising starting point towards larger-scale quantum systems based on colour-center qubits. 

\section*{Results}

\textbf{System: NV electron-nitrogen spin qubits.}
We consider a single NV center in a type-IIa isotopically purified diamond at 4 K. We study the two-qubit register formed by the NV electron spin and the nitrogen ($^{14}$N) nuclear spin, and consider the surrounding $^{13}$C nuclear spins (concentration $\sim 0.01 \%$) and P1 center electron spins (concentration $\sim 75$ ppb) as sources of noise (Fig. 1a) \cite{Degen2021,Bradley2022}.
The Hamiltonian of this system is \cite{Doherty2012}:
\begin{equation}\label{H_sys}
\begin{split}
H = DS_{z}^{2} + \gamma_{e} B_{z}S_{z} + \gamma_{e} B_{\perp}S_{x} + QI_{z}^{2}+\gamma_{n}B_{z}I_{z} \\
 + \gamma_{n}B_{\perp}I_{x} + A_{xx}S_{x}I_{x} + A_{yy}S_{y}I_{y} + A_{zz}S_{z}I_{z},
\end{split}
\end{equation}
where $D \approx 2.87$ GHz is the zero-field splitting of the electron spin, $\gamma_{e} = 2.8024$ MHz/G ($\gamma_{n} = -307.7$ Hz/G) is the gyromagnetic ratio of the electron (nitrogen) spin, $Q \approx -4.949$ MHz is the quadrupole splitting of the nitrogen spin and $[A_{xx},A_{yy},A_{zz}] \approx [2.68, 2.68, 2.188]$ MHz are the diagonal hyperfine components of the spin-spin interaction between the electron spin and the nitrogen spin (Supplementary Note II). $[S_x, S_y, S_z]$ $([I_x, I_y, I_z])$ are the spin-1 operators for the electron (nitrogen) spin. We work at an approximately aligned, external magnetic field of $B_z \approx 62.29$ G and $B_{\perp} \approx 0.41$ G (Supplementary Note V). We choose two levels of each qutrit to encode a qubit: $m_{s}=\{0,-1\}$ and $m_I = \{0, -1\}$ for the electron and nitrogen spin respectively (Fig. 1b).

\begin{figure}
\includegraphics[width=\columnwidth]{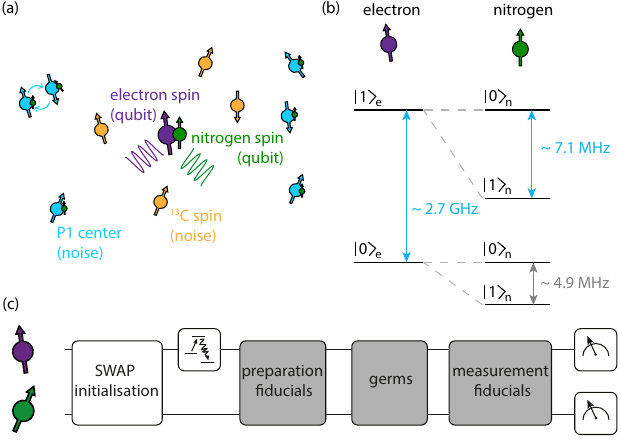}
\caption{\textbf{System and experiment overview.} \textbf{a.} The system under consideration is the NV center in diamond. The electron spin-1 (purple) and $^{14}$N spin-1 (green) are surrounded by two spin baths that generate noise: nuclear $^{13}$C spins and P1-center electron spins. \textbf{b.} Schematic level diagram of the electron and nitrogen spin. The transitions indicated in blue are used for the electron and nitrogen gates. \textbf{c.} Experimental implementation of gate set tomography. We initialise the electron spin optically and the nitrogen spin via a SWAP with the electron spin (Supplementary Note VIII). Then, we run a set of preparation circuits (fiducials), germs and measurement circuits, after which the electron and nitrogen spin are read out. The germs are longer sequences of gates meant to amplify specific types of errors \cite{Nielsen2021,Blume-Kohout2022}.}
\label{fig1}
\end{figure}

We initialise and read out the NV electron spin through resonant, optical excitation \cite{Robledo2011}. Before each experimental repetition, we prepare the NV center in the negative charge state and the lasers on resonance with the NV transitions \cite{Bernien2013}. To initialise the nitrogen spin, we swap the electron $m_s = 0$ state to the nitrogen in a two-step SWAP initialisation process (Supplementary Note VIII). To read out the nitrogen spin, we map its spin state on the electron spin, which is then read out optically (Supplementary Note IX). \\

\textbf{Gate set tomography (GST).}
To characterise and optimise gates, we use gate set tomography (GST). GST is a calibration-free tomography method used to characterise an informationally complete set of quantum gates \cite{Nielsen2021,Blume-Kohout2022}. It has previously been applied on a variety of systems, such as ion traps \cite{Blume-Kohout2017,Proctor2020}, quantum dots \cite{Xue2022}, silicon donors \cite{Madzik2022,Dehollain2016} and superconducting qubits \cite{White2021}. The process entails executing a set of quantum circuits, consisting of preparation and measurement circuits (called fiducials) and germs that are used to amplify specific types of gate errors (Fig. 1c). Compared to benchmarking methods, such as randomized benchmarking \cite{Emerson2005,Magesan2011,Helsen2022,Wallman2014,Proctor2017,Epstein2014}, interleaved randomized benchmarking \cite{Magesan2012,Sheldon2016,Helsen2022} and cross-entropy benchmarking \cite{Boixo2018,Arute2019}, GST estimates the complete process matrices for the gate set. In this work, we exploit that detailed information of the gate errors to systematically characterise and optimise gates for the electron-nitrogen two-qubit system. \\ 

\begin{figure*}
\includegraphics[width=\textwidth]{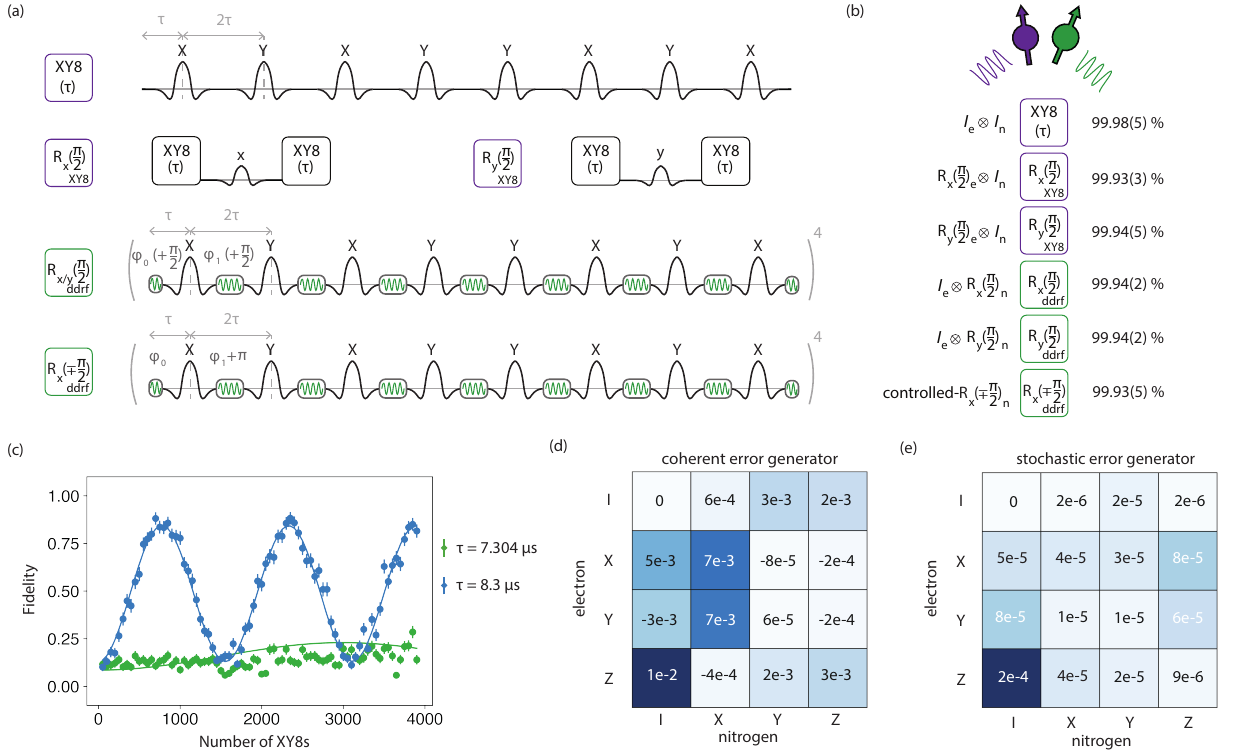}
\caption{\textbf{Two-qubit GST on the electron-nitrogen system.} \textbf{a.} Pulse sequences for the electron, nitrogen and two-qubit gates. The identity operation is implemented by an XY8 decoupling sequence. The electron $\pi/2$ pulses are placed between XY8 sequences. The nitrogen single-qubit gates are implemented using DDRF \cite{Bradley2019}. For the two-qubit gate, all odd RF pulses get an additional $\pi$ phase shift, which makes the nitrogen rotation conditional on the electron-spin state. \textbf{b.} Average gate fidelities for the electron-nitrogen two-qubit system (see Methods for the definition of average gate fidelity). For each operation (left), we show the gate implementation (middle, see (a)) and the average gate fidelity (right). The single-qubit gate fidelities take into account the effect on the other qubit (see Methods). The error bars represent one standard deviation (a $67 \%$ confidence interval). \textbf{c.} Effect of the XY8 dynamical-decoupling sequence ($I_e \otimes I_n$) of (a) on the nitrogen spin for different interpulse delays $\tau$. We initialise the nitrogen spin in $m_I = 0$ and apply a variable number of XY8 sequences with $\tau = 7.304$ \textmugreek s (green) or $\tau = 8.3$ \textmugreek s (blue). We find a coherent rotation of the nitrogen spin at $\tau = 8.3$ \textmugreek s whereas this is absent for $\tau = 7.304$ \textmugreek s. The blue line for $\tau = 8.3$ \textmugreek s is a fit that yields the perpendicular magnetic field $B_\perp \approx 0.41$ G (Supplementary Note V). The green line for $\tau = 7.304$ \textmugreek s is a simulation of the expected signal given the extracted $B_\perp$. The error bars on the data represent one standard deviation. \textbf{d,e.} Coherent and stochastic error generators for the controlled gate ($R_x(\mp\pi/2)$). The box colours are proportional to the magnitude of the indicated values. Dephasing of the electron spin (stochastic, $Z\otimes I$) is the main contribution to the average gate infidelity.}
\label{fig2}
\end{figure*}

\textbf{Complete two-qubit gate set.}
A universal set of quantum gates requires a two-qubit gate, as well as single-qubit gates for each qubit. Note that, in general, one is interested in the complete evolution within the two-qubit space. That is, a single-qubit rotation on one of the qubits should simultaneously realise the identity operation on the other qubit. For example, a single-qubit $\pi/2$ gate on the nitrogen nuclear spin is defined as $I_e \otimes R_x(\pi/2)_n$, with $I_e$ the identity matrix. 
  
A key challenge is that the electron- and nitrogen-spin qubits have an `always-on' hyperfine interaction (Eqn. \ref{H_sys}), and that the electron spin continuously couples to the surrounding spin bath. Therefore, a central element of this work is to design composite gates that can precisely decouple unwanted interactions during the gate operations. 

In specific cases, the conditions for single-qubit gates can be relaxed. For example, when at a given moment in an algorithm the other qubit is in a known computational eigenstate or its information does not need to be preserved, some interactions do not have to be decoupled. Below, we first consider the general case of the full two-qubit space. We discuss single-qubit gates optimised for specific cases in a later section.  

As basic operations on the electron-spin qubit we use $\pi$ and $\pi/2$ pulses, which we implement by applying Hermite pulse shapes modulated by the electron-spin frequency ($\sim 2.7$ GHz). A Hermite pulse shape is chosen to minimize the effects of the detunings introduced by the different nitrogen-spin states (Supplementary Note VI). 

Out of these $\pi$ and $\pi/2$ pulses, we construct composite $\pi/2$ and identity gates (Fig. 2a). All gates include dynamical decoupling of the electron spin in order to avoid phase errors due to the electron-nitrogen interaction (predominantly of the form $S_zI_z$, see equation \ref{H_sys}), and to decouple from the environment. To optimise the electron gates, we calibrate the interpulse delay $\tau$, as well as the amplitude of the electron $\pi$ and $\pi/2$ pulses for both the gates around $x$ and $y$ (Supplementary Note XII). 

To implement gates involving the nitrogen spin, we use dynamical decoupling radio-frequency (DDRF) gates \cite{Bradley2019}. The DDRF gate consists of dynamical decoupling on the electron spin interspersed with RF pulses ($\sim 7.1$ MHz) on the nitrogen-spin qubit (Fig. 2a). This decouples the electron spin from its environment while simultaneously driving the nitrogen spin. By updating the phases between consequent RF pulses to account for the nitrogen-spin precession, an effective rotation builds up. By shifting the overall phase of the RF pulses, we can rotate the nitrogen spin around either $x$ or $y$.

This rotation can be chosen to be unconditional (a single-qubit gate) or conditional on the electron-spin state (a two-qubit gate), by setting the phases of the RF pulses \cite{Bradley2019}. This conditional interaction is a controlled $\mp \pi/2$ gate around $x$:
\begin{equation}
    \ket{0}\bra{0}_e \otimes R_x(-\pi/2)_n + \ket{1}\bra{1}_e \otimes R_x(\pi/2)_n, 
\end{equation}
where $0$ ($1$) refers to the electron-spin qubit being in $m_s = 0$ ($m_s = -1$). To optimise the nitrogen gates, we apply multiple, consecutive $\pi/2$ gates, while sweeping the RF amplitude of the DDRF gates (Supplementary Note XII). No separate optimisation is performed for the two-qubit gate.

After optimisation (see below for an extended discussion), we observe single-qubit gate fidelities around $99.94 \%$ and a two-qubit gate fidelity of $99.93(5) \%$ for full two-qubit operation (Fig. 2b, see Methods for the definition of average gate fidelity). These fidelities are amongst the highest reported in any system \cite{Scholl2023,Xue2022,Noiri2022,Sung2021,Kandala2021,Clark2021,Ballance2016}. Compared to a previous characterisation of different NV-center gates using subspace randomized benchmarking \cite{Xie2023}, the fidelities we obtain are best-estimates rather than upper limits, we obtain the full process matrices, and we implement a complete gate set that includes nitrogen-spin single-qubit gates. 

To investigate the stability of the system and the robustness of the calibration routines, we perform a total of four two-qubit GST experiments. Across all experiments, we consistently find fidelities above $99.9 \%$ for all gates. The two-qubit gate fidelity averaged over the four runs is $F_{avg} = 99.923 \pm 0.026 \%$ (Supplementary Note XIX). \\

\textbf{Optimising the interpulse delay $\tau$.}
An important gate optimisation step is to set the interpulse delay $\tau$ of the electron XY8 decoupling sequence. Ideally, this sequence cancels the interaction of the electron-spin qubit with the spin environment and with the nitrogen-spin qubit, performing an identity gate on both qubits. However, such decoupling sequences can resonantly couple the NV electron-spin qubit to the dynamics of the surrounding spin bath, which includes the dynamics of individual nuclear spins \cite{Taminiau2012,Kolkowitz2012,Abobeih2018,Bradley2022}, as well as of nuclear-spin pairs \cite{Bartling2022} and electron-spin pairs \cite{Bartling2023}. We leverage our detailed knowledge of the nuclear- and electron-spin environment of this NV center \cite{Degen2021,Bradley2022,Bartling2023} to choose suitable values of the interpulse delay $\tau$, at which we avoid decoherence of the electron spin due to the $^{13}$C and P1 spin baths (Supplementary Note V).

Next, we optimise the electron XY8 decoupling sequence for performing the identity operation on the nitrogen-spin qubit. To remove residual $z$-rotations we set $\tau$ to a multiple of the period set by the average precession frequency of the nitrogen spin for the two electron states ($\tau = 4 \pi n / (\omega_0 + \omega_{-1}$)).

However, for this value of $\tau$, we observe a small but significant unconditional coherent rotation on the nitrogen-spin qubit for the electron XY8 decoupling sequence (Fig. 2c). This coherent rotation can be explained by an effective $S_z I_x$ interaction \cite{Taminiau2012,Liu2019} between the electron and nitrogen spin coming from the misaligned magnetic field of $B_{\perp} \approx 0.41$ G, which breaks the symmetry of the system (Supplementary Note V) \cite{Liu2019}. 

To remove this nitrogen-spin rotation, we impose a second condition: $\tau = 2 \pi m / (\omega_0 - \omega_{-1})$. Together with the first condition, this ensures that a single dynamical decoupling unit $(\tau - \pi - 2 \tau - \pi - \tau)$ does not induce any effective rotation on the nitrogen-spin qubit, independent of the exact field alignment (Fig. 2c, Supplementary Note V). \\

\textbf{Gate errors.}
GST provides completely-positive trace-preserving process matrices for the gate set \cite{Nielsen2021,Blume-Kohout2022}. Since the process matrix offers a complete description of the two-qubit evolution under the action of a gate, it provides information about the type, origin and magnitude of infidelities in our system. Error generators can be used to dissect the process matrices in distinct error sources with clear physical interpretations \cite{Blume-Kohout2022}. In particular, the coherent Hamiltonian and the incoherent stochastic error generators distinguish between unitary error processes, such as over- and under-rotations, and stochastic error processes, such as qubit dephasing. 

For the two-qubit gate, the largest remaining error source is the electron single-qubit coherent error around $z$ (Fig. \ref{fig2}d). However, coherent errors only contribute quadratically to the average gate fidelity, whereas stochastic errors contribute linearly \cite{Nielsen2002,Sanders2016,Blume-Kohout2017}. Therefore, we conclude that the average gate fidelities of the gates in this work are mainly limited by single-qubit dephasing ($Z_e I_n$) and single-qubit bit flips ($Y_e I_n$) of the electron-spin qubit (Supplementary Note XXI). 

The electron-spin dephasing and bit flips likely originate from magnetic field noise. The main candidates for the source are the fluctuating external magnetic field and the spin environment of the NV center (Fig. 1a). The time-scale of the noise plays an important role in its effect on the gates. Simulations show that quasi-static noise, which fluctuates slowly and can be considered as constant during a single gate, has no significant effect on the gate fidelity due to the XY8 decoupling sequences (Supplementary Note VII). However, higher frequency noise, which fluctuates within a single gate, may explain the observed electron-spin dephasing and bit flips. A likely source are the spin-bath dynamics, in particular flip-flops ($\sim$ kHz) of interacting P1 centers ($\sim 75$ ppb) \cite{Degen2021,Bartling2023}. \\

\textbf{Context-specific single-qubit gates.} For both single- and two-qubit gates, the two-qubit process matrix is the most complete description (Fig. 2). However, gates optimised using two-qubit process matrices are not necessarily optimal under all circumstances. For example, if one qubit is in a well-defined state (e.g. an eigenstate) or does not hold valuable information and is allowed to dephase, then more optimal or simpler single-qubit gate designs likely exist for the other qubit. Such situations are very common in quantum algorithms, for example in quantum error correction where ancillary qubits are regularly measured and re-initialized \cite{Abobeih2018}. In what follows, we consider various single-qubit gate designs and their performance in different cases. To characterise these gates, we perform GST on the single-qubit system of either the electron-spin or nitrogen-spin qubit.

\begin{figure}
\includegraphics[width=\columnwidth]{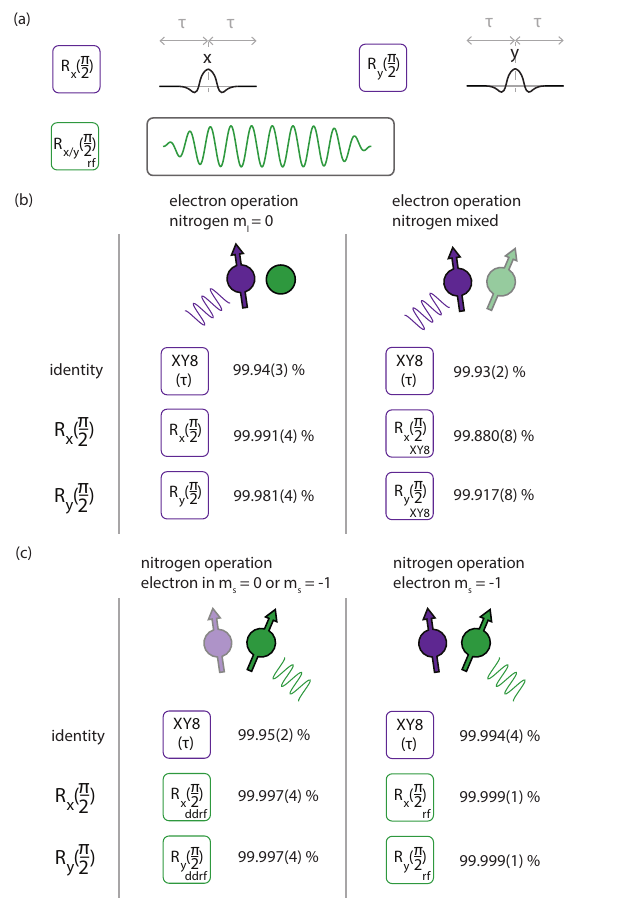}
\caption{\textbf{Single-qubit GST for the electron-spin and nitrogen-spin qubits.} \textbf{a.} Pulse sequences. We additionally implement single-qubit gates without XY8 decoupling for the electron-spin qubit and for the nitrogen-spin qubit (RF pulses; length $\sim 100$ \textmugreek s; risetime $1$ \textmugreek s). \textbf{b.} Average gate fidelities from single-qubit GST for the electron-spin qubit for two different cases. (left) The nitrogen spin is in $m_I = 0$ and the $\pi/2$-gates do not include XY8 decoupling. (right) The nitrogen spin is mixed and the $\pi/2$-gates include XY8 decoupling. \textbf{c.} Average gate fidelities from single-qubit GST for the nitrogen-spin qubit, for two cases. (left) The $\pi/2$-gates are implemented using DDRF, which can be used regardless of the electron-spin state. Here, the electron spin is in $m_s = 0$. In Supplementary Note XVIII we show the result for $m_s = -1$, obtaining similar fidelities. (right) The $\pi/2$-gates are implemented using simple RF pulses (no decoupling), with the electron-spin state being $m_s = -1$. For b and c, the error bars represent one standard deviation (a $67 \%$ confidence interval).}
\label{fig3}
\end{figure}

\begin{figure*}
\includegraphics[width=\textwidth]{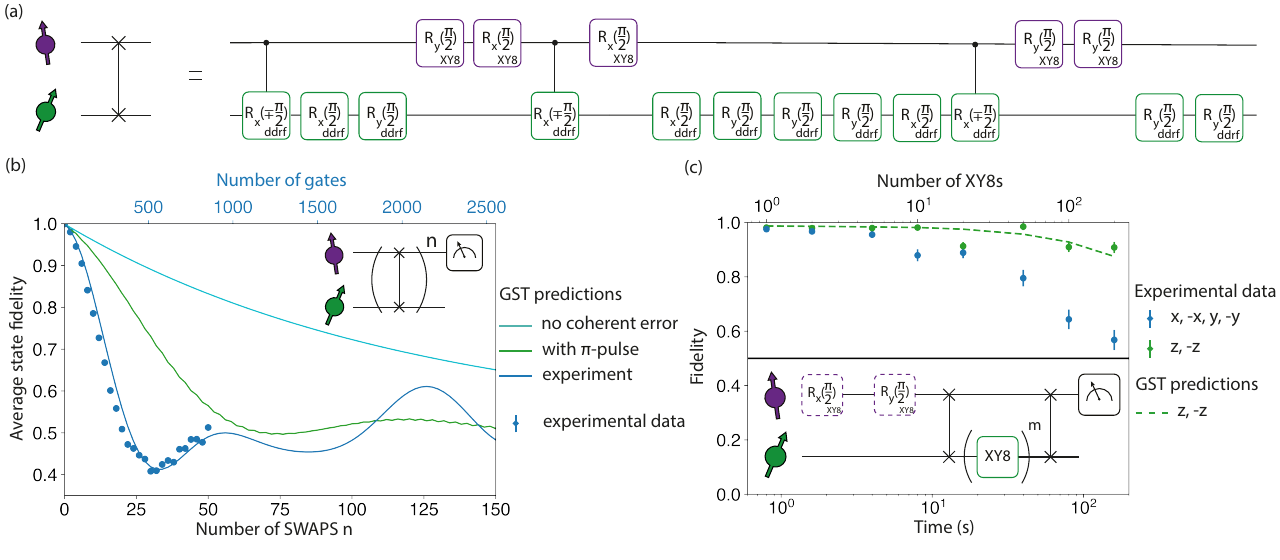}
\caption{\textbf{Implementation of an electron-nitrogen SWAP gate.} \textbf{a.} Sequence to perform an electron-nitrogen SWAP gate composed of the gates characterised by two-qubit GST (Fig. 2). The compiled sequence is not generally optimal, but demonstrates the utility of GST in predicting the action of extended circuits. \textbf{b.} Application of $n$ sequential SWAP gates. The experimental data (blue points) are predicted well by the GST process matrices (blue line). The GST predictions without any coherent error (cyan) and with an additional $\pi$-pulse between sequential SWAP gates (green) are also shown. The error bars are smaller than the data points. \textbf{c.} Storage of all six electron cardinal states ($\pm x, \pm y, \pm z$) on the nitrogen spin with the SWAP gate, showing the protection of a quantum state for over $100$ s. We compare the experimental data for $\pm z$ to a GST prediction that does not include any free evolution. We find that the $\pm z$ decay is well explained by the obtained process matrix and thus likely caused by gate imperfections. The nitrogen XY8 decoupling is implemented with the DDRF $\pi/2$ gates around $x$ and $y$ characterised by two-qubit GST (Fig. 2). The error bars on the data represent one standard deviation.}
\label{fig4}
\end{figure*}

We first consider single-qubit gates on the electron spin for the specific case in which the nitrogen spin is in a known eigenstate. In this case, it is not necessary to decouple the hyperfine coupling, so that a gate consisting of a simple microwave pulse can be used (Fig. \ref{fig3}a). Fig. \ref{fig3}b (left) shows the gate fidelities from GST for $m_I = 0$. Compared to the gate designs with decoupling (Supplementary Note XVIII), we find a slight improvement in fidelity, which we attribute to the removal of the XY8 decoupling sequences. Additionally, the direct microwave pulse is much faster ($1.344$ \textmugreek s vs. $234.728$ \textmugreek s) and reduces the stochastic gate errors, so that it is preferred when the nitrogen spin is in a known eigenstate during algorithms.

Next, we consider the case where the nitrogen spin is fully mixed ($m_I = 0, \pm 1$). This case applies when the nitrogen spin is not (yet) used in an algorithm, which is common, for example, during sensing experiments or remote entanglement generation \cite{Pompili2021,Hensen2015,Hermans2022}. In this case, the strong hyperfine coupling makes it preferable to apply decoupling, and the simple microwave pulse (Fig. \ref{fig3}a) shows significantly lower fidelity ($\sim 99.86\%$, Supplementary Note XVIII). Fig. \ref{fig3}b (right) shows that the fidelities for the gates with decoupling are slightly reduced compared to the pure nitrogen-spin state, as the hyperfine interaction ($A_{zz} \approx 2.188$ MHz) combined with the limited peak Rabi frequency ($\Omega \sim 27$ MHz) cause different electron-spin rotations for the three nitrogen states (Supplementary Note VI).

For the nitrogen-spin qubit, we first characterise the DDRF-based single-qubit gates (Fig. 2a) in the single-qubit subspace for different electron-spin states (Fig. \ref{fig3}c and Supplementary Note XVIII). We find fidelities of $99.997(4) \%$, confirming that the gate fidelity in the two-qubit subspace is mainly limited by errors on the electron spin. When the electron is in a known eigenstate (here $m_s = -1$), the nitrogen-spin gates can alternatively be implemented by a simple radio-frequency (RF) pulse (Fig. 3a) at the corresponding frequency ($\sim 7.1$ MHz), without decoupling. We introduce a risetime of $1$ \textmugreek s and additionally we match the pulse length ($\sim 100$ \textmugreek s) to be a multiple of the nitrogen-spin precession frequency to avoid any effective $z$-rotation. For this case, we observe fidelities of $99.999(1) \%$ (Fig. 3c), limited by the measurement uncertainty. These gates outperform the DDRF-based single-qubit gates, for the specific case where the electron spin is in an eigenstate, and the fidelities obtained are among the highest single-qubit gate fidelities reported on any platform \cite{Ballance2016,Harty2014,Leu2023,Barends2014,Rong2015,Xue2022,Noiri2022,Yang2019,Yoneda2018}.

To explicitly demonstrate a universal gate set, we also implement a T-gate (or $\pi/4$-gate) for both the electron-spin and nitrogen-spin qubit using the gate designs that include XY8 decoupling (as in Fig. 2a). We characterise the T-gates in the single-qubit subspace (Supplementary Note XVIII). For the electron-spin qubit, we find an average gate fidelity of $F = 99.989(7) \%$, and for the nitrogen-spin qubit we find $F = 99.986(6) \%$. 

For comparison, we also perform single-qubit randomized benchmarking \cite{Emerson2005,Helsen2022} on the electron $\pi/2$-gates with and without XY8 decoupling, as well as on the nitrogen DDRF $\pi/2$-gates. We compare the average fidelity from randomized benchmarking to simulations based on the process matrices obtained from GST (Supplementary Note XIV). We generally find good agreement between the two methods, with small deviations likely originating from the presence of non-Markovian noise \cite{Blume-Kohout2017}. A detailed analysis is given in Supplementary Note XIV. \\

\textbf{SWAP gate implementation.} The two-qubit process matrices from GST provide a complete characterisation, which can be used to predict the outcome of any gate sequence or algorithm on the two qubits. In principle, for the two-qubit system, no other gate sequences need to be investigated. Nevertheless, it is instructive to consider the compilation of extended circuits and algorithms from the characterised gate set. In particular, the effect of coherent errors (Fig. 2d) and their suppression by error mitigation methods depends on the specific sequences of gates in an algorithm.

As an example, we choose to implement an electron-nitrogen SWAP gate and use it to swap quantum states from and to the nitrogen-spin qubit, which provides a long-lived quantum memory. We construct a single SWAP gate using a total of 17 gates from the characterised gate set (Fig. 4a). Note that a SWAP operation can be compiled with fewer gates, but this choice allows us to validate the predictive power of the obtained GST process matrices. From the two-qubit process matrices, we obtain an expected SWAP gate fidelity of $98.7 \%$. 

To investigate the SWAP gate, we repeatedly apply it, swapping a quantum state between the two qubits (Fig. 4b). We apply the SWAP sequence an even number of times with the electron spin prepared along each of its cardinal axes. After about 20 SWAP gates, the resulting average state fidelity reaches $0.5$. The prediction using the process matrices obtained with GST is in good agreement with the experimental results, even up to $\sim 800$ elementary gates (Supplementary Note XV), indicating that the GST model and its assumptions are accurate.

Next, we theoretically analyse the limitations of the SWAP gate using the error generators (Fig. 2d,e). The main source of the observed decay are coherent errors that build up by repeatedly applying the SWAP gate. Upon theoretically excluding the coherent errors from the process matrices, a much slower decay is predicted, going well beyond $150$ SWAP gates (Fig. 4b). Removing all coherent errors is not easily achieved experimentally. However, based on the explicit knowledge of the coherent error processes, tailored sequences might be designed to cancel their effect. A simple example is to add an inversion $\pi$-gate (two extra electron $\pi/2$ gates) between every SWAP gate, which cancels part of the build-up of coherent errors (Fig. 4b). Alternatively, error mitigation techniques such as Pauli twirling \cite{Knill2005,Emerson2005,Kern2005,Wallman2016} can be used to prevent coherent errors from adding up (see the analysis in Supplementary Note XVI). 

The example in Fig. 4b underlines the limitation of using a single number as a metric of the quality of a gate. The average gate fidelity captures the effect of decoherence during the gates properly. However, if unitary errors build up or cancel in a compiled circuit, it becomes impossible to predict the error rate from the average gate fidelity alone \cite{Sanders2016}. Process matrices obtained with GST, on the other hand, predict the behaviour of such compiled circuits accurately and can be used to design optimized sub-routines (Supplementary Note XV). 

Finally, as an example of a potential application for the SWAP gate, we use the nitrogen-spin qubit as a quantum memory to temporarily store an arbitrary quantum state. First, we prepare the electron spin in one of six cardinal states and swap the state to the nitrogen-spin qubit. Then, we apply XY8 dynamical decoupling on the nitrogen-spin qubit using the GST-characterised DDRF $\pi/2$ gates. We measure the average state fidelity of the nitrogen spin as a function of the number of XY8 decoupling sequences (Fig. 4c). We find that the quantum state on the nitrogen spin can be maintained for over $100$ s, showing its promise as a quantum memory (see also Ref. \cite{Bradley2019}). 

We compare the experimental data with a simulation based on the two-qubit GST process matrices (without the inclusion of a free evolution time). From this simulation, we find that the errors in the nitrogen DDRF $\pi/2$ gates can explain the decay of the nitrogen $\pm z$ eigenstates (Fig. 4c). Additionally, we find that the electron-spin population decays at a timescale similar to the decoherence of the nitrogen $\pm x$ and $\pm y$ states (Supplementary Note XVII). This suggests that the current measurement is limited by gate imperfections that affect the electron-spin population, and that longer coherence times are likely achievable by using the gates of Fig. 3, but this is not pursued here.

\section*{Discussion}

In conclusion, we have designed, characterised, and systematically optimised a universal set of single- and two-qubit gates on the electron-nitrogen spin system of the NV center using gate set tomography. A central element of the gate design and optimisation is to precisely decouple the (unwanted) effects of the electron-nitrogen hyperfine interaction. We have demonstrated single-qubit gate fidelities of up to $99.999(1)\%$ and a two-qubit gate fidelity of $99.93(5) \%$. These are among the most accurate quantum gates shown on any platform \cite{Ballance2016,Harty2014,Leu2023,Barends2014,Rong2015,Xue2022,Noiri2022,Yang2019,Yoneda2018,Xie2023,Scholl2023,Sung2021,Kandala2021,Clark2021}, and present a step towards enabling larger-scale algorithms. 

Future work might further improve the gates by targeting the dominant error generators identified here (Fig. 2d,e). Additionally, the methods developed here provide a starting point for optimizing gates for a variety of colour centers in various materials, including diamond, silicon carbide and silicon \cite{Nagy2019,Pingault2017,Rosenthal2023}, as well as for gates in multi-qubit systems \cite{Bradley2019,Abobeih2022}. 

\section*{Methods}

\textbf{Experimental setup and sample.} The experiments are performed on a type-IIa isotopically purified (targeted $0.01 \%$ $^{13}$C) $\langle 100 \rangle$ diamond substrate (Element Six). We use a home-built confocal microscope to address a single NV center at 4 K. A solid immersion lens and anti-reflection coating are fabricated around the NV center to increase the collection efficiency \cite{Bernien2013}. We use three orthogonal permanent NdFeB magnets mounted on linear actuators (Newport UTS100PP) to apply an external magnetic field aligned along the NV symmetry axis. To initialise and read out the NV electron-spin state, we use resonant optical excitation (Toptica DLPro and New Focus TLB-6704-P). We measure readout fidelities of $83.3(4) \%$ for the $m_s = 0$ spin state and $98.9(1) \%$ for the $m_s = -1$ spin state, obtaining an average readout fidelity of $91.1(2) \%$. To prepare the NV in the NV$^{-}$ charge state and lasers on-resonance with the NV$^{-}$ transitions, we perform a charge-resonance check \cite{Bernien2013}, which additionally involves 515 nm (green, Cobolt MLB) excitation. Through direct current modulation or cascaded acousto-optical modulators, we realize on/off ratios exceeding 100 dB for all lasers so that the electron spin relaxation $T_1$ is negligible \cite{Abobeih2018}.

\textbf{Microwave and RF driving.} For electron-spin driving, we use single-sideband modulation at 250 MHz. The I \& Q signals are generated on an arbitrary waveform generator (Tektronix AWG5014C), which modulates an RF source (R\&S SGS100A). We use a 20 W amplifier (AR 20S1G4) to attain peak Rabi frequencies of $\sim 27$ MHz, followed by a microwave switch, which is shut when microwaves are not applied to protect the NV from amplifier noise \cite{Abobeih2019}. For nitrogen-spin driving, we generate the RF frequency directly from the arbitrary waveform generator. Finally, the electron-spin and nitrogen-spin driving signals are combined on a diplexer. For more details on the electronics, see Supplementary Note I.

\textbf{Magnetic field stabilisation.} To mitigate magnetic field fluctuations, we intermittently (every $\sim 10-20$ min.) calibrate the electron-spin resonance frequency to within $2$ kHz of the set point by moving one of the three permanent magnets. Due to this intermittent calibration, the typical peak-to-peak fluctuations of the electron-spin resonance frequency during the two-qubit GST experiments do not exceed $10$ kHz.

\textbf{Data analysis.} Here, we summarise the experimental settings and model violation for the presented GST results. We group the results per figure and indicate the corresponding name in that figure (Table \ref{gst_settings_main}). First, we show the maximum circuit depth $L$ used for that experiment. When we indicate $L = 128$, that implies that germs of depth $1, 2, 4, 8, 16, 32, 64$ and $128$ were run \cite{Nielsen2021}. Then, we indicate the experimental repetitions used for each circuit. The estimation error of gates in GST is typically of the order $O(1/(L\sqrt{N}))$ where $N$ is the number of repetitions \cite{Nielsen2021}. To perform the GST analysis, we use pyGSTi \cite{pygsti}. For the process matrix, we require complete positivity and trace preservation (CPTP). Lastly, we report the $N_\sigma$ metric for $L=128$, which quantifies the model violation \cite{Nielsen2021}. The error bar on the average gate fidelities obtained with GST represent one standard deviation (a $67 \%$ confidence interval). Other important elements of the GST experiments are the preparation fiducials, germs and measurement fiducials used. We provide these separately, together with the data.

\textbf{Average gate fidelity.} In this work, we use the average gate fidelity as a metric to summarize the quality of each gate. The average gate fidelity is calculated by comparing the process matrix in the Pauli transfer matrix representation \cite{Chow2012}, obtained through gate set tomography, to the ideal target process matrix \cite{Nielsen2002}:
\begin{equation}\label{avg_fid}
\begin{split}
F_{\text{avg}} = \frac{\text{Tr}(P^{\dagger}_{\text{exp}}P_{\text{target}})/d+1}{d+1}
\end{split}
\end{equation}
where $P_{\text{exp}}$ ($P_{\text{target}}$) is the process matrix for the experimental (ideal) gate in the Pauli transfer matrix representation such as \cite{Nielsen2002}
\begin{equation}\label{avg_fid}
\begin{split}
(P_{\Lambda})_{ij}=\frac{1}{d} Tr(\sigma_{i} \Lambda(\sigma_{j}))
\end{split}
\end{equation}
where $\Lambda$ is the map of the gate to be characterized,  $\sigma_{i(j)}$ is the Pauli operator for axis $i$ ($j$) and $d$ is the dimension of the system of interest. 

In this work, we primarily consider two-qubit GST and process matrices of dimensions $16 \cross 16$ ($d=4$, two-qubit space) (Figs. 2 and 4). In the section ``Context-specific single-qubit gates'' we use single-qubit GST, which only considers a single-qubit subspace, and thus process matrices of dimensions $4 \cross 4$ ($d=2$, single-qubit space) (Fig. 3). In the latter case, deviations from the identity process on the other qubit do not directly result in a reduced gate fidelity. 

\begin{table}[t]
\begin{tabular}{|c|c|c|c|} 
 \hline
  \multicolumn{4}{|c|}{Fig. 2} \\ \hline
name & L & repetitions & $N_\sigma$ \\ \hline 
full operation & 128 & 1000 & 11.7 \\ \hline \hline
  \multicolumn{4}{|c|}{Fig. 3} \\ \hline
name & L & repetitions & $N_\sigma$ \\ \hline 
 electron operation nitrogen $m_I = 0$ & 128 & 1000 & 8.71 \\ \hline
 electron operation nitrogen mixed & 128 & 1000 & 15.4 \\ \hline
    nitrogen operation electron $0$ or $-1$ & 128 & 2000 & 1.3 \\ \hline
  nitrogen operation electron $m_s = -1$ & 512 & 1000 & 2.76 \\ \hline \hline
\end{tabular}
\caption{\textbf{Settings and model violation for the presented GST experiments.} $L$ is the maximum circuit depth, repetitions is the number of experimental repetitions per circuit and $N_\sigma$ quantifies the model violation \cite{Nielsen2021}.}
\label{gst_settings_main}
\end{table}

\bibliography{bib.bib}

\section*{Acknowledgements}
We thank R.J. Blume-Kohout, C.I. Ostrove, J. Gonzalez de Mendoza, M. Rimbach-Russ, J. Borregaard for valuable discussions. 
We gratefully acknowledge support from the joint research program “Modular quantum computers” by Fujitsu Limited and Delft University of Technology, co-funded by the Netherlands Enterprise Agency under project number PPS2007. This work was supported by the Netherlands Organisation for Scientific Research (NWO/OCW) through a Vidi grant, as part of the Frontiers of Nanoscience (NanoFront) programme and through the project “QuTech Phase II funding: "Quantum Technology for Computing and Communication” (Project No. 601.QT.001). This project has received funding from the European Research Council (ERC) under the European Union’s Horizon 2020 research and innovation programme (grant agreement No. 852410). This work was supported by the Dutch National Growth Fund (NGF), as part of the Quantum Delta NL programme. This work is part of the research programme NWA-ORC (NWA.1292.19.194), which is partly financed by the Dutch Research Council (NWO). This research was supported by the education and training program of the Quantum Information Research Support Center, funded through the National research foundation of Korea (NRF) by the Ministry of Science and ICT (MSIT) of the Korean government (No. 2021M3H3A103657313). The fitting algorithms were performed on the DelftBlue supercomputer at Delft University of Technology \cite{DHPC2022}.

\section*{Author contributions}
HPB, JY and THT devised the experiments. HPB, JY, KNS and MvR performed the experiments and collected the data. HPB, JY, KNS, MvR, LAE and HBvO prepared the experimental apparatus. HPB, JY, KNS, MvR, LAE, HBvO, MB, FS and THT analyzed the data. MM and DJT grew the diamond sample. HPB, JY, KNS, MvR and THT wrote the manuscript with input from all authors. MB, FS and THT supervised the project.

\clearpage

\end{document}


\begin{center}
    {\Large \centering \bf Supplementary Information}
\end{center}

\tableofcontents

\clearpage

\section{Experimental setup (microwave, radio-frequency)} \label{mw_setup}
\begin{figure*}[h]
\includegraphics[width=\textwidth]{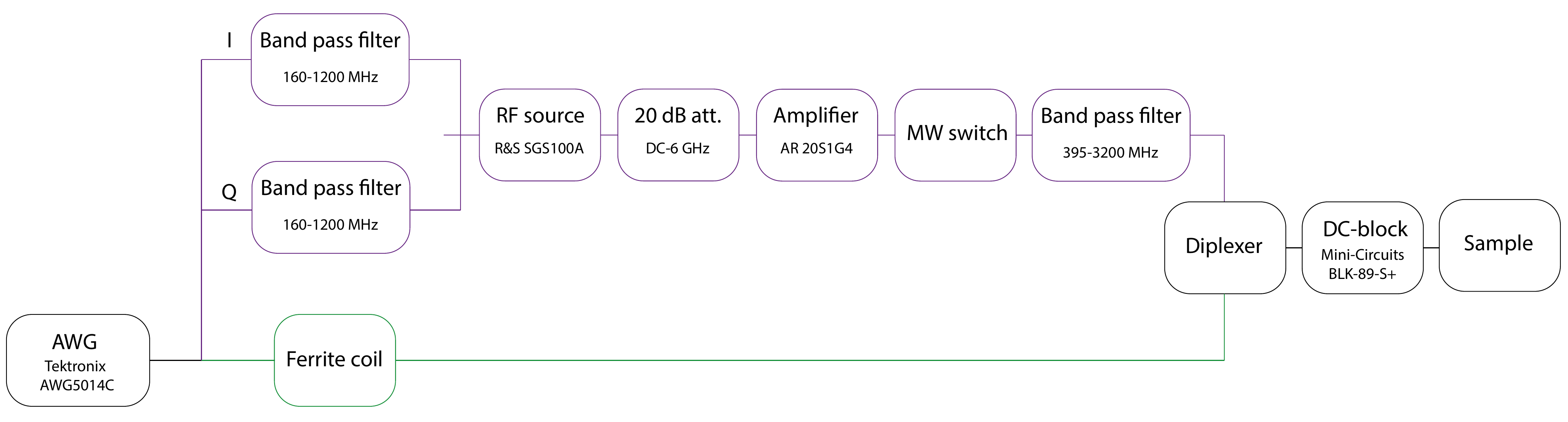}
\caption{\textbf{Diagram of the microwave and radio-frequency delivery system.} For the electron spin (purple), we generate single-sideband modulated pulses at $250$ MHz, which are fed into the RF source. We attenuate by 20 dB to protect the amplifier. The MW switch aims to reduce noise when no pulses are applied. Band pass filters are added to filter low-frequency noise originating from the AWG or MW switch. For the nitrogen spin (green), we synthesize the pulses directly from the AWG. A ferrite coil is used to suppress high-frequency electronic noise. The electron and nitrogen drives are combined on a diplexer, fed through a DC block and then to the sample.}
\label{figs10}
\end{figure*}

In this section, we discuss the experimental setup to drive electron microwave pulses and nitrogen radio-frequency pulses. In Fig. \ref{figs10} we show a schematic containing all relevant components. 

For the electron-spin pulses, we apply microwave pulses resonant to the electron energy-level splitting ($\sim 2.7$ GHz) with a Hermite pulse envelope \cite{Warren1984}. We additionally use single-sideband modulation. We set the signal generator (R\&S SGS100A) to a detuned frequency ($\sim 2.7$ GHz $+250$ MHz), and generate a waveform with a frequency of $250$ MHz with the arbitrary waveform generator (AWG, Textronix AWG5014C). A specific phase relation between the I and Q channel leaves only the lower sideband ($\sim 2.7$ GHz), which is resonant to the electron energy-level splitting. This single-sideband modulation is advantageous to filter out the inherent low-frequency noise of the AWG by adding band-pass filters after the AWG output.

The signal is then passed through an RF amplifier (AR 20S1G4), followed by a home-built microwave switch. This switch is designed to pass the RF signal only when a certain above-threshold voltage is present, which is controlled by the AWG. It is optimized such that the signal chain is open only when microwave pulses are applied.

The $\sim 7.1$ MHz frequency to drive the nitrogen spin is synthesized directly from the AWG. We add a ferrite coil to remove high-frequency electronic noise. The MHz nitrogen frequency and GHz electron frequency are combined on a home-built diplexer, after which it passes through a DC block (Mini-Circuits BLK-89-S+) and goes to the sample which is located in a Montana Cryostation S50. To get the microwave and radio-frequency signals close to the NV center, we pattern a gold stripline on the diamond surface.

\clearpage

\section{Hamiltonian of the system} \label{hamiltonian}
\begin{figure*}[h]
\includegraphics[width=0.7\textwidth]{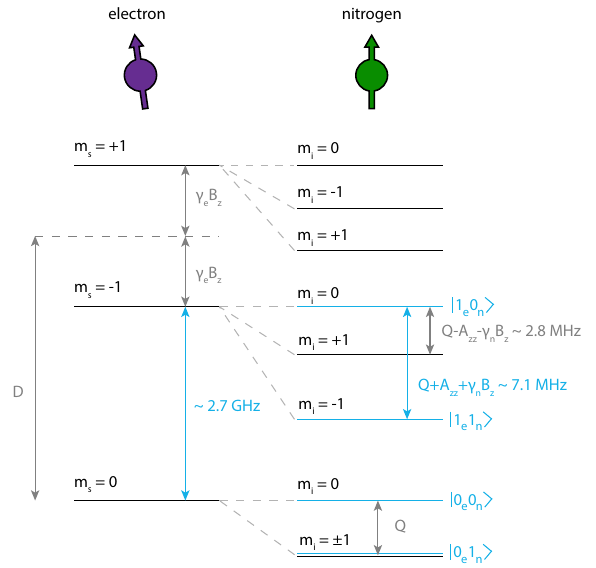}
\caption{\textbf{Level diagram of the electron-nitrogen spin-system.} $m_s$ ($m_I$) is the electron (nitrogen) spin, $\gamma_e$ ($\gamma_n$) is the electron (nitrogen) gyromagnetic ratio, $D = 2.873668(9)$ GHz is the electron zero-field splitting, $Q = -4.949156(1)$ MHz is the nitrogen quadrupole splitting, and $A_{zz} = 2.188218(2)$ MHz is the $zz$-component of the electron-nitrogen hyperfine interaction. The qubit levels used in this work are indicated by $\ket{0_e 0_n}$, $\ket{0_e 1_n}$, $\ket{1_e 0_n}$ and $\ket{1_e 1_n}$, where $e$ ($n$) stands for the electron (nitrogen) spin. The transitions used for driving in the quantum gates are indicated in blue, and the $\sim 2.8$ MHz nitrogen transition is used for SWAP initialisation of the $^{14}$N spin (Section \ref{nitrogen_init}).} 
\label{en_levels}
\end{figure*}

In this work, we investigate an NV center in an isotopically purified diamond (targeted $0.01 \%$ $^{13}$C). The Hamiltonian of this system can be well approximated by an isolated two-spin system \cite{Doherty2012}:
\begin{equation}\label{H_sys}
H = DS_{z}^{2} + \gamma_{e} B_{z}S_{z} + \gamma_{e} B_{\perp}S_{x} + QI_{z}^{2} + \gamma_{n}B_{z}I_{z}+ \gamma_{n}B_{\perp}I_{x} + A_{xx}S_{x}I_{x} + A_{yy}S_{y}I_{y} + A_{zz}S_{z}I_{z} 
\end{equation}
where $[S_x, S_y, S_z]$ $([I_x, I_y, I_z])$ are the Pauli spin-1 operators for the electron (nitrogen) spin, $D \approx 2.874$ GHz is the zero-field splitting of the electron spin, $\gamma_{e}$ ($\gamma_n$) is the electron (nitrogen) gyromagnetic ratio, $B_{z}$ ($B_{\perp}$) is the external magnetic field parallel (perpendicular) to the NV axis, $Q$ is the quadrupole splitting of the nitrogen spin, and $[A_{xx}, A_{yy}, A_{zz}]$ are the diagonal hyperfine components of the spin-spin interaction between the electron spin and the nitrogen spin. Note that this Hamiltonian does not include the couplings to P1 centers ($\sim 75$ ppb, order $\sim$ kHz) and $^{13}$C nuclear spins ($\sim 0.01 \%$, order $\sim 100$ Hz), which provide a source of noise in this work \cite{Degen2021, Bradley2022, Bartling2023}.  In Figure \ref{en_levels}, we show a level diagram of the electron-nitrogen spin system.

To obtain the coefficients for the Hamiltonian of our system, we measure 6 resonance frequencies: the nitrogen $m_I = 0$ to $m_I = \pm 1$ transition for the two different electron states $m_{s} = 0$ and $m_{s} = -1$, and the two electron $m_{s} = 0$ to $m_{s} = -1$ transitions for the two different nitrogen states $m_{I} = 0$ and $m_{I} = -1$. These values correspond to the eigenenergy differences between the corresponding energy levels. The eigenenergies can be calculated by diagonalizing the Hamiltonian of the system. We can approximate the diagonalized Hamiltonian with time-independent perturbation theory up to first-order using the diagonal components of the system Hamiltonian as the non-perturbed Hamiltonian and using all of the off-diagonal components as the perturbation. This is a valid assumption since the off-diagonal components ($A_\perp$, $B_\perp$) are more than 3 orders-of-magnitude smaller than the difference between the diagonal components. In this framework, we obtain six approximate equations, where we have excluded terms smaller than $A_{\perp}^{2}/D$:

\begin{equation}\label{wi_m1_msm1}
\omega(m_I: 0 \leftrightarrow -1|m_{s}=-1) = -Q + \gamma_{n} B_{z} + A_{zz} + A_{\perp}^{2}/D = 7.120706(1) \text{ MHz}
\end{equation}
\begin{equation}\label{wi_p1_msm1}
\omega(m_I: 0 \leftrightarrow +1|m_{s}=-1) = -Q - \gamma_{n} B_{z} - A_{zz} = 2.780105(6) \text{ MHz}
\end{equation}
\begin{equation}\label{wi_m1_ms0}
\omega(m_I: 0 \leftrightarrow +1|m_{s}=0) = -Q - \gamma_{n} B_{z} - A_{\perp}^{2}/D = 4.965825(3) \text{ MHz}
\end{equation}
\begin{equation}\label{wi_p1_ms0}
\omega(m_I: 0 \leftrightarrow -1|m_{s}=0) = -Q + \gamma_{n} B_{z} - A_{\perp}^{2}/D = 4.927491(1) \text{ MHz}
\end{equation}
\begin{equation}\label{ws_p1_ms0}
\omega(m_s: 0 \leftrightarrow -1|m_{I}=-1) = D - \gamma_{e}B_{z}+A_{zz}+A_{\perp}^{2}/D = 2.701294(1) \text{ GHz}
\end{equation}
\begin{equation}\label{wi_p1_ms0}
\omega(m_s: 0 \leftrightarrow -1|m_{I}=0) = D - \gamma_{e}B_{z}+3A_{\perp}^{2}/D = 2.699101(1) \text{ GHz}
\end{equation}

where $A_{\perp} = A_{xx} = A_{yy}$. These six frequencies can be obtained by measuring the resonance lines from optically detected magnetic resonance (ODMR) and Ramsey measurements. 

From these equations, the coefficients of the Hamiltonian can be derived, which are $D = 2.873668(9)$ GHz, $B_{z} = 62.291(3)$ G, $Q = -4.949156(1)$ MHz, $A_{zz} = 2.188218(2)$ MHz, and $A_{\perp} = 2.679(1)$ MHz. The confidence range on the coefficients are propagated from the error range on the frequency values measured from the ODMR and Ramsey measurements. The gyromagnetic ratio of the electron spin is $\gamma_{e} = 2.8024$ MHz/G and that of the nitrogen spin is $\gamma_{n} = -307.7$ Hz/G. The magnetic field misalignment ($B_{\perp} \approx 0.414$ G) that will be discussed later (section VI) has a small effect on the coefficients of the two-spin system compared to $A_\perp^2/D$. For the simulations throughout this work, we use these values to represent our system.

\clearpage

\section{Data analysis}
In this section, we discuss the interpretation of the uncertainties on the parameters obtained from our fits. In general, the objective of a least-squares fit is to minimize the goodness-of-fit parameter $\chi^2$ \cite{Bevington2003}:

\begin{equation}
    \chi^2 = \sum_i \Big[\frac{y_i-y(x_i)}{\sigma_i} \Big]^2,
\end{equation}
where the sum is over the experimental data points, $y_i$ is the experimental data, $\sigma_i$ the experimental error bar, and $y(x_i)$ the prediction of the fit for $x_i$. Ideally, the deviation of the data from the fit matches the error bars on the data points, resulting in $\chi^2 = N-M$, where $N$ is the number of data points and $M$ the number of fit parameters. We can define the reduced goodness-of-fit parameter as \cite{Bevington2003}

\begin{equation}
    \chi_r^2 = \chi^2/\text{DOF},
\end{equation}
where $\text{DOF} = N-M$ is the number of degrees of freedom of the fit.

Obtaining $\chi_r^2 = 1$ implies that the deviation of the data from the fit is well explained by the error bars on the data $\sigma$. However, if $\chi_r^2 \gg 1$ ($\chi_r^2 \ll 1$), the data error bars predict significantly smaller (larger) differences between the data and the model. This points to under-estimated (over-estimated) noise on the data, statistical fluctuation due to small amounts of data, or a need for developing further refinements of the theoretical model to fit. 

The common approach is to rescale the covariance matrix by $\chi_r^2$ \cite{Bevington2003}. This is the default method in Python packages such as \verb|lmfit| and \verb|scipy|, and the values we report follow this approach. Alternatively, the covariance matrix is not rescaled and this results in different (larger or smaller, depending on $\chi_r^2$) error bars on the fit parameters. Where deemed important, we additionally report $\chi_r^2$ in order to enable easy calculation of the error without rescaling.

\section{Simulation details}\label{sim}
In this work, we developed a simulation tool based on QuTiP \cite{Johansson2012,Johansson2013} to simulate the electron-nitrogen spin system. The simulation assumes that there are no other spins. For the ideal two-spin system, we use the Hamiltonian described in equation \ref{H_sys}. The simulation of the unitary operation of the gates was done using the propagator function in QuTiP. To also consider non-rotating frame effects, we solve the master equation with the full time-dependent Hamiltonian to obtain the unitary representation of our gates. The simulation results in Fig. 2c and Figs. S3c,d, S4, S5 are based on this tool.

\clearpage

\section{XY8 sequence: decoupling the electron-nuclear interaction} \label{B_field}
Our gate designs use dynamical decoupling sequences to decouple the interaction between the electron and nuclear spin, as well as the interactions of the electron spin with the surrounding spin baths and other noise sources. We then realize single- and two-qubit gates involving the $^{14}$N nuclear spin by adding direct RF driving. 

\subsection{Decoupling from the $^{13}$C and P1-center baths}
During dynamical decoupling, the NV electron-spin qubit can couple to single nuclear-spin dynamics \cite{Taminiau2012,Kolkowitz2012,Abobeih2018,Bradley2022}, as well as electron- \cite{Bartling2023} and nuclear-spin \cite{Bartling2022} pair dynamics. In particular, we have previously demonstrated the control of a single $^{13}$C spin \cite{Bradley2022} and of a single pair of P1 centers \cite{Bartling2023} surrounding this NV center. To avoid decoherence of the NV electron-spin qubit due to these nuclear and electron spins, we choose a suitable value of the interpulse delay $\tau$. There is no observable interaction between the NV electron spin and the P1-center pair for dynamical decoupling sequences below $\tau = 10$ \textmugreek s \cite{Bartling2023}. Additionally, at an external magnetic field of $B_z = 62.291(3)$ G, the interpulse delays $\tau$ at which the $^{13}$C nuclear spins (concentration $\sim 0.01 \%$) couple coherently to the NV electron-spin qubit are $\tau = (2k+1) \tau_0$ where $\tau_0 = 1/(4 \gamma_c B_z) \approx 3.75$ \textmugreek s. Together with the two other conditions (see below), this leads us to choose $\tau = 7.304$ \textmugreek s, at which $\tau$ we avoid coupling to the nuclear-spin bath and electron-spin P1-pairs.

\subsection{Decoupling the electron-nitrogen interaction}
Above, we considered the decoupling of the NV electron-spin qubit from the environment. However, we also need to consider unwanted evolution of the nitrogen-spin qubit during the electron XY8 decoupling sequences, introduced by the hyperfine interaction. To minimize this, we satisfy two additional conditions for the interpulse delay $\tau$ in dynamical decoupling. 

First, we set $\tau$ to a multiple of the period given by the average precession frequency of the nitrogen spin in $m_s = 0$ and $m_s = -1$. Second, we set $\tau$ to a multiple of the period defined by the effective electron-nitrogen hyperfine interaction. That is, if $\omega_0$ ($\omega_{-1}$) is the precession frequency of the nitrogen spin when the electron spin is in $m_s = 0$ ($m_s = -1$), we set $ \tau = 4 \pi n/ (\omega_0 + \omega_{-1})$ where $n$ is an integer and simultaneously $ \tau = 2 \pi m/ (\omega_0 - \omega_{-1})$ where $m$ is an integer.

\subsubsection{The first condition: $ \tau = 4 \pi n/ (\omega_0 + \omega_{-1})$} \label{first_tau_condition}
The first condition ensures that the nitrogen spin undergoes a multiple of a $2 \pi$-rotation around $z$ for each gate. In this way, we minimize control errors around $z$. Due to a hardware constraint, $\tau$ is limited to multiples of $4$ ns, which is relatively easy to satisfy when the average precession period is already a multiple of $2$ ns. Therefore, we set the magnetic field such that the average precession frequency of the nitrogen spin is a multiple of $2$ ns. 

We adjust the magnetic field using the two electron-spin resonance frequencies: the transitions from $m_s = 0$ to $m_s = \pm 1$. This is an efficient way to approximately align the magnetic field by minimizing the sum of the $m_s = 0$ to $m_s = \pm 1$ transition frequencies \cite{Balasubramanian2008}. Additionally, we move to a magnetic field magnitude that satisfies the first condition described above: we set the magnetic field such that the period of the average nitrogen precession frequency is 166 ns (a multiple of 2 ns).

\subsubsection{Nitrogen-spin rotation for $\tau = 4 \pi n/ (\omega_0 + \omega_{-1})$}
After satisfying the first condition, we observe a coherent rotation of the nitrogen nuclear spin under dynamical decoupling of the electron spin (Fig. \ref{figs11}). This rotation of the nitrogen nuclear spin has been observed previously in the context of DC field sensing \cite{Liu2019}. The slightly misaligned magnetic field in our system breaks the symmetry of the hyperfine interaction and induces an effective $A_{zx} S_z I_x$ interaction between the electron and nitrogen spin. This term then introduces a coherent rotation on the nitrogen spin when applying a dynamical decoupling sequence with a $\tau$ that satisfies the first condition, as is known from for example the control of $^{13}$C nuclear spins \cite{taminiau2014}. Note that the effective $A_{zx}$ coupling of the nitrogen under misaligned magnetic field was discussed in depth by Liu et al. \cite{Liu2019}. For completeness, we summarise the related discussion here. The effective $A_{zx}$ originating from a misaligned magnetic field is: 

\begin{align} \label{Azx}
    A_{zx} = \frac{\gamma_{e}B_{\perp}A_{\perp} }{D-\gamma_{e}B_{z}}F
\end{align} 
where $B_{\perp}$ is the off-axis magnetic field and $F \approx \frac{2 \sqrt{2}(-Q+A_{zz}/2)^{2} }{(-Q+A_{zz})Q}$. These equations assume that the external magnetic field is such that the NV center is close to the ground-state level-anticrossing (GSLAC), which is not satisfied in our work, rendering some approximations used in Ref. \cite{Liu2019} quantitatively inaccurate. Compared to an NV close to the GSLAC (see Ref. \cite{Liu2019}), the nitrogen-spin rotation in this work is a more subtle effect that only appears when applying on the order of 1000 XY8 sequences (8000 $\pi$-pulses).

Under the presence of such an $A_{zx}$ term, we can predict the rotation angle $\phi$ of the nitrogen spin per single dynamical decoupling unit consisting of ($\tau - \pi - 2\tau - \pi - \tau$) using the following equation \cite{Taminiau2012}:

\begin{align} \label{nitrogen_rotation_frequency}
    \cos(\phi)=\cos(\alpha)\cos(\beta)-\cos(\theta)\sin(\alpha)\sin(\beta)
\end{align}

where $\alpha = \omega_{-1} \tau$, $\beta = \omega_{0} \tau$ and $\theta$ is the angle between the two rotation axes of the nitrogen spin for the two different electron eigenstates ($m_s=0$ and $m_s=-1$). The two rotation axes are obtained from the exact eigenstates we extract from the system Hamiltonian in equation \ref{H_sys}.

\subsubsection{The second condition: $ \tau = 2 \pi m/ (\omega_0 - \omega_{-1})$}
To avoid this unwanted rotation on the nitrogen-spin qubit, we require $\phi = 0$. To that end, either $\alpha = 0 \text{ (mod } 2\pi)$ and $\beta = 0 \text{ (mod } 2\pi)$ or $\alpha = \pi \text{ (mod } 2\pi)$ and $\beta = \pi \text{ (mod } 2\pi)$. In other words, we require $\alpha - \beta = 0 \text{ (mod } 2\pi)$. This can be rewritten to obtain the second condition outlined before: $\tau = 2 \pi m/ (\omega_0 - \omega_{-1})$.

Note that we can rewrite the two conditions ($\tau = 4 \pi n/ (\omega_0 + \omega_{-1})$ and $\tau = 2 \pi m/ (\omega_0 - \omega_{-1})$) as:

\begin{align} \label{tau_conditions}
    2 \tau &= \frac{2\pi (2 n + m)}{\omega_0}, \\
    2 \tau &= \frac{2\pi (2 n - m)}{\omega_{-1}}.
\end{align}

There are two ways to satisfy these equations. The first type of solution is if $\tau$ is an integer multiple of the periods set by both $\omega_0$ and $\omega_{-1}$. This solution is obtained if $2n+m$ (and therefore $2n-m$) is an even number. Intuitively, in that case the evolution during a single $\tau$ equals the identity operation for both electron states and no nitrogen spin rotation can be created. For the second type of solution, $2n+m$ is odd (and therefore $2n-m$ is odd). In that case, the evolution during a single $\tau$ is an effective $\pi$-evolution, for both $m_s = 0$ and $m_s = -1$, which also becomes the identity operation for a full unit of the dynamic decoupling sequence ($\tau-\pi-2\tau-\pi-\tau$).

Because setting these conditions results in $\phi=0$, the $^{14}$N evolution becomes the identity operation, independent of the magnetic field alignment. This choice of parameters thus makes the decoupling sequence --- and by extension the gates in this work --- intrinsically robust to magnetic field misalignment.

\subsection{Calibrating the magnetic field to minimise the nitrogen-spin rotation}
The rotation of the nitrogen spin during XY8 decoupling was measured for two different values of $\tau$: $\tau = 7.304$ \textmugreek s and $\tau = 8.3$ \textmugreek s. At the magnetic field obtained to satisfy the first condition for $\tau$ (Sec. \ref{first_tau_condition}), $\tau = 7.304$ \textmugreek s is already a close multiple of the effective electron-nitrogen interaction time, while $\tau = 8.3$ \textmugreek s is not. 

To measure the leftover rotation of the nitrogen spin at these values of $\tau$, we apply many XY8 dynamical decoupling sequences with the nitrogen starting in $m_I = 0$ (Fig. \ref{figs11}). For a suboptimal external magnetic field magnitude, we find that the nitrogen spin is rotated from $m_I = 0$ to $m_I = -1$ by an XY8 decoupling sequence at $\tau = 7.304$ \textmugreek s (Fig. \ref{figs11}a). Additionally, the nitrogen rotation present at $\tau = 8.3$ \textmugreek s does not show full contrast (Fig. \ref{figs11}b). We optimize the magnetic field magnitude such that no nitrogen rotation is visible when decoupling at $\tau = 7.304$ \textmugreek s (Fig. \ref{figs11}c) and a full-contrast nitrogen rotation is visible when decoupling at $\tau = 8.3$ \textmugreek s (Fig. \ref{figs11}d). This is equivalent to optimising the magnetic field such that $\tau = 7.304$ \textmugreek s satisfies both conditions: $ \tau = 4 \pi n/ (\omega_0 + \omega_{-1})$ and $ \tau = 2 \pi m/ (\omega_0 - \omega_{-1})$ with $n=44$ and $m=8$. This magnetic field ensures that no spurious nitrogen rotations are introduced at $\tau = 7.304$ \textmugreek s by the XY8 decoupling. In Fig. \ref{figs11}c we verify that for our magnetic field we see no visible rotation of the nitrogen spin up to 4,000 XY8 sequences on the electron spin (32,000 $\pi$-pulses). 

\subsection{Magnitude of the magnetic field misalignment}
\begin{figure*}[h]
\includegraphics[width=0.8\textwidth]{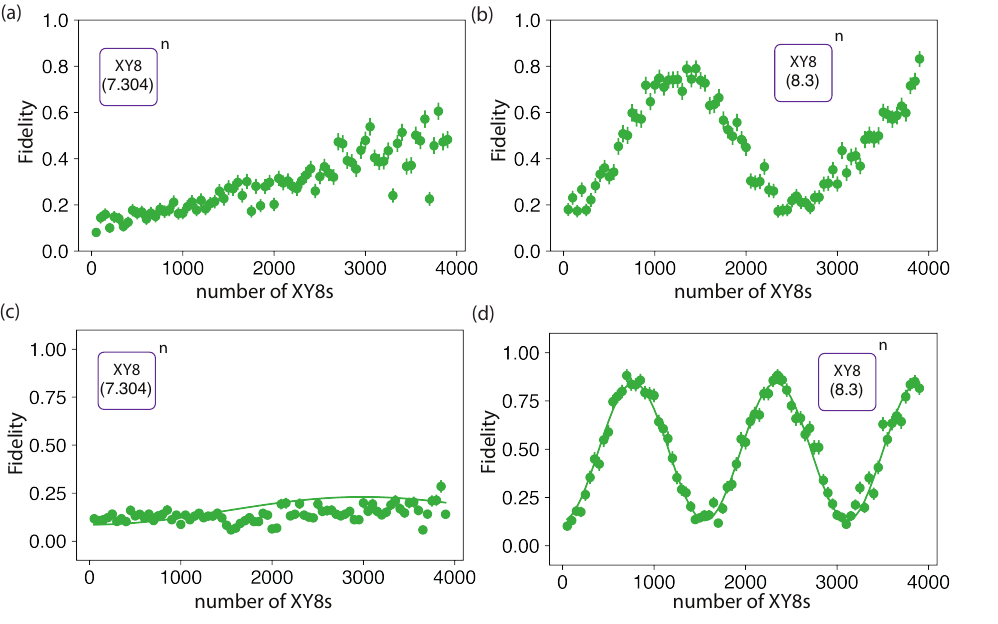}
\caption{\textbf{Nitrogen rotation under electron XY8 decoupling for two different magnetic fields.} \textbf{a.} The nitrogen spin is initialised in $m_I = 0$ and under electron XY8 decoupling at $\tau = 7.304$ \textmugreek s, we observe a slow rotation of the nitrogen spin. This indicates that the condition $ \tau = 2 \pi m/ (\omega_0 - \omega_{-1})$ is not exactly matched for $\tau = 7.304$ \textmugreek s. \textbf{b.} Nitrogen rotation under electron XY8 decoupling at $\tau = 8.3$ \textmugreek s. The reduced contrast compared to (d) indicates that the condition $ \tau = 4 \pi n/ (\omega_0 + \omega_{-1})$ is not exactly matched for $\tau = 8.3$ \textmugreek s and thus for $\tau = 7.304$ \textmugreek s. \textbf{c.} Nitrogen rotation under electron XY8 decoupling at $\tau = 7.304$ \textmugreek s for an optimal external magnetic field. No rotation of the nitrogen spin under XY8 decoupling is observed, since $\tau = 7.304$ \textmugreek s satisfies both conditions at this magnetic field: $ \tau = 4 \pi n/ (\omega_0 + \omega_{-1})$ and $ \tau = 2 \pi m/ (\omega_0 - \omega_{-1})$. \textbf{d.} For the optimal magnetic field setting, we observe a full-contrast rotation under decoupling at $\tau = 8.3$ \textmugreek s. This happens because $\tau = 8.3$ \textmugreek s is not a multiple of the effective electron-nitrogen interaction time ($ \tau \neq 2 \pi m/ (\omega_0 - \omega_{-1})$), while it is a multiple of the period set by the average nitrogen-spin precession frequency ($ \tau = 4 \pi n/ (\omega_0 + \omega_{-1})$). We fit the data in d to a simulation of XY8 decoupling on the system Hamiltonian, with $B_\perp$ as a free parameter (solid line). We find $B_\perp \approx 0.414$ G, which we then use to simulate the data in c (solid line).}
\label{figs11}
\end{figure*}

To investigate the required perpendicular magnetic field to reproduce this effect, we simulate the action of XY8 decoupling sequences on the electron-nitrogen spin system. We take the system Hamiltonian in equation \ref{H_sys} and simulate the effect of each $\pi$ pulse using a time-dependent Hamiltonian propagator calculation as described in section \ref{sim}. In Figure \ref{figs11}c,d, we show the simulation result for the nitrogen spin when a perpendicular magnetic field is present and when $\tau$ is $100/(\omega_{0}+\omega_{1}) \approx 8.3 $ \textmugreek s and $88/(\omega_{0}+\omega_{1}) \approx 7.304 $ \textmugreek s. We fit to the experimental result for $\tau = 8.3$ \textmugreek s in order to obtain $B_{\perp}$. We find $B_{\perp} = 0.41400000(1)$ G with a reduced chi-squared of the fit of $\chi^2_r = 1.8$. Note that the confidence range of the perpendicular field is from the error of the fit, and the uncertainties on the coefficients within the Hamiltonian used in the model were not considered during the fit.

The simulations in Figure \ref{figs11} together with the data presented in Figure 2 and Figure \ref{figs11} show that the conditions outlined above are important to avoid any coherent rotation of the nitrogen spin. In future experiments, the necessity of the first condition ($ \tau = 4 \pi n/ (\omega_0 + \omega_{-1})$) can be avoided through the use of more advanced phase-tracking methods.

\clearpage

\section{Gate designs} \label{gates}
\begin{figure*}[h]
\includegraphics[width=\textwidth]{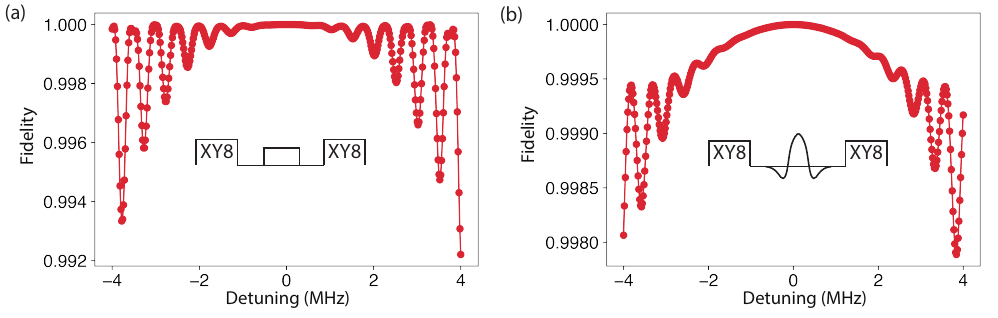}
\caption{\textbf{Pulse fidelity of a square and Hermite $\pi/2$ pulse as a function of detuning.} \textbf{a.} Square $\pi/2$ pulse fidelity as a function of microwave frequency detuning. We find a rapid deterioration of pulse fidelity in the presence of detuning. We sandwich the square $\pi/2$ pulse in between XY8 decoupling sequences made of square $\pi$ pulses. \textbf{b.} Hermite $\pi/2$ pulse fidelity as a function of microwave frequency detuning. We obtain high pulse fidelities for a much larger range of detunings compared to the square pulse. For both the square pulse and Hermite pulse, we use a maximum Rabi frequency of $12.693$ MHz for the $\pi/2$ pulse. We sandwich the Hermite $\pi/2$ pulse in between XY8 decoupling sequences made of Hermite $\pi$ pulses with a maximum Rabi frequency of $26.653$ MHz.}
\label{figs12}
\end{figure*}

In this section, we discuss some details of the electron and nitrogen pulses. We use a microwave frequency resonant with the $m_s = 0$ to $m_s = -1$ transition when $m_I = 0$. The main issue for the electron single-qubit gate is the relatively large interaction strength between the electron and nitrogen nuclear spin. Since the detuning induced by the coupling of the nitrogen spin ($\sim 2$ MHz) is comparable to our peak Rabi frequency ($\sim 27$ MHz), we use a Hermite envelope generated with the following formula:
\begin{align} \label{Hermite}
    f(t) = A\Bigg(1-\eta \Big(\frac{t-T_{pulse}/2}{0.1667 T_{pulse}}\Big)^{2}\Bigg) \exp{-\Big(\frac{t-T_{pulse}/2}{0.1667 T_{pulse}}\Big)^{2}}
\end{align}

where $A$ is the amplitude of the Hermite envelope, $\eta$ is a coefficient that is $0.956$ ($0.667$) for a $\pi$ ($\pi/2$) pulse and $T_{pulse}$ is the defined pulse length (in this work, $T_{pulse} =144$ ns). \cite{Warren1984,Vandersypen2005}. This makes it possible to rotate the electron spin similarly regardless of the nitrogen spin state. We can write the Hamiltonian of the NV electron spin for a given nitrogen eigenstate $m_I$ as
\begin{equation}\label{H_e}
H_{e} =  DS_{z}^{2} + \gamma_{e} BS_{z} + m_{I} A_{zz}S_{z},
\end{equation}
where $A_{zz} \approx 2.188$ MHz is the electron-nitrogen hyperfine interaction. If we now add a time-dependent Hamiltonian rotating at $D + \gamma_{e}B$ and move into the rotating wave frame of this frequency, the Hamiltonian after applying rotating wave approximation is:
\begin{equation}\label{H_e_rwa}
H_{e} =  m_{I} A_{zz}S_{z} + \Omega_{e} (\cos(\phi)S_{x}+\sin(\phi)S_y). 
\end{equation}
Thus the electron has a different detuning depending on the nitrogen-spin state. Therefore, when applying the same microwave pulse, the effective rotation is along a different rotation axis and has a different angle for different $m_I$. The Hermite envelope for the microwave pulse helps the phase accumulation during the microwave pulse to be more robust to detuning \cite{Warren1984,Vandersypen2005}. Figure \ref{figs12}a (\ref{figs12}a) shows a simulation of the fidelity of a Hermite (square) $\pi/2$ pulse sandwiched between XY8 decoupling sequences as a function of the microwave frequency detuning. For both  $\pi/2$ pulses, we use the same maximum Rabi frequency of $12.693$ MHz and for the $\pi$ pulses in the XY8 sequence, we use the same maximum Rabi frequency of $26.653$ MHz. For the Hermite pulse envelope, a gate with fidelity over $99.9 \%$ over a range of $\pm 2$ MHz of detuning is possible, whereas for a square envelope the fidelity drops and fluctuates rapidly as a function of detuning outside the $\pm 1$ MHz range. 

The nitrogen-spin gates consist of decoupling sequences on the electron spin, as described above, combined with radio-frequency pulses. Alternatively, we apply only radio-frequency pulses of $\sim 100$ \textmugreek s in duration with the electron in $m_s = -1$ (Fig. 3). We apply the radio-frequency pulses resonant with the $m_I = 0, -1$ transition for $m_s = -1$ at a frequency of 7.120706 MHz. We add an error function envelope to each RF pulse with a risetime of 1 \textmugreek s. 

\clearpage

\section{Effect of quasi-static noise on XY8 decoupling}
\begin{figure*}[h]
\includegraphics[width=0.5\textwidth]{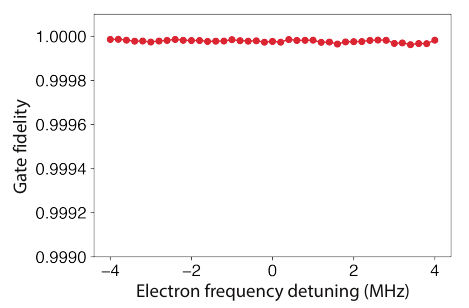}
\caption{\textbf{Simulation of the effect of electron frequency detuning on the average gate fidelity of XY8 decoupling.} We simulate the application of a single XY8 decoupling block as a function of the electron frequency detuning. We find that the effect of detuning is negligible.}
\label{figs13}
\end{figure*}

In this section, we discuss the effect of quasi-static magnetic noise on the electron XY8 decoupling sequence. Such noise can originate from magnetic field fluctuations, either from the externally applied magnetic field or from the spin bath ($^{13}$C nuclear spins and P1 center electron spins). For quasi-static noise, we take the magnetic field as constant for the duration of a gate. For almost all gates that we discuss, we use a decoupling sequence on the electron spin. This makes sure that any phase picked up by the electron spin due to the quasi-static bath is cancelled out. Figure \ref{figs13} shows a simulation of how an XY8 sequence (identity gate) responds to a constant magnetic field detuning by extracting the average gate fidelity of the XY8 as an identity gate from the process matrix of an XY8 sequence in the electron single-qubit subspace. The simulation was conducted with the full two-qubit Hamiltonian of the system, so the gate fidelity contains information on how both the electron and nitrogen react to the quasi-static environment. From this result, we see that for a quasi-static environment, the XY8 sequence (identity gate) should ideally not show any additional infidelity.

\clearpage

\section{Nitrogen-spin initialisation} \label{nitrogen_init}

\begin{figure*}[h]
\includegraphics[width=\textwidth]{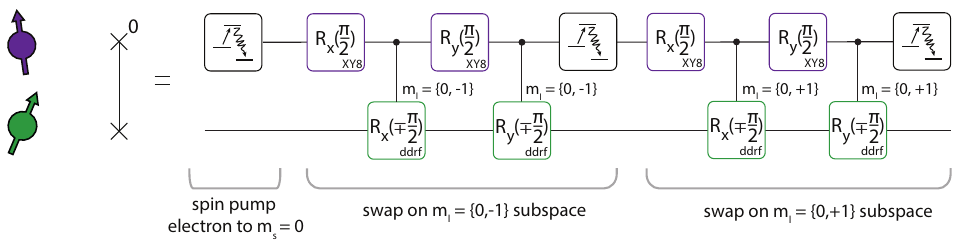}
\caption{\textbf{Experimental sequence for SWAP initialisation of the nitrogen spin.} We perform a two-step SWAP process due to the spin-1 nature of the nitrogen spin. First, we perform a swap on the $m_I = \{0,-1\}$ subspace, after which we perform a swap on the $m_I = \{0,+1\}$ subspace. The gates are as defined in the main text. However, the gates on the subspace $m_I = \{0, +1\}$ of the nitrogen spin utilise a different RF frequency (Supplementary Section \ref{additional_gst_results}). For the SWAP icon on the left, the superscript $0$ indicates that this is not a full SWAP (c.f. Fig. 4), but that the initialisation of the electron spin in $m_s = 0$ is a requirement.}
\label{figs3}
\end{figure*}

In this section, we discuss the initialisation of the nitrogen spin to its $m_I = 0$ state. In the gate set tomography results presented in the main text, we use SWAP-type initialisation of the nitrogen spin. In Fig. \ref{figs3} we show the corresponding sequence. We perform a SWAP between the electron $m_s = \{0,-1\}$ subspace and the nitrogen spin. Since the nitrogen spin is a spin-1 system, and we swap qubit-to-qubit, we need a two-step SWAP process. First, we initialise the electron spin in $m_s = 0$. Then, we perform a swap on the $m_I = \{0,-1\}$ subspace of the nitrogen spin. Next, we reinitialise the electron spin after which we perform a swap on the $m_I = \{0,+1\}$ subspace. Note that we use a reduced type of SWAP gate: as the electron spin is set to an eigenstate, the first of the three two-qubit gates can be omitted compared to the full SWAP gate of Fig. 4.

\begin{figure*}[h]
\includegraphics[width=0.5\textwidth]{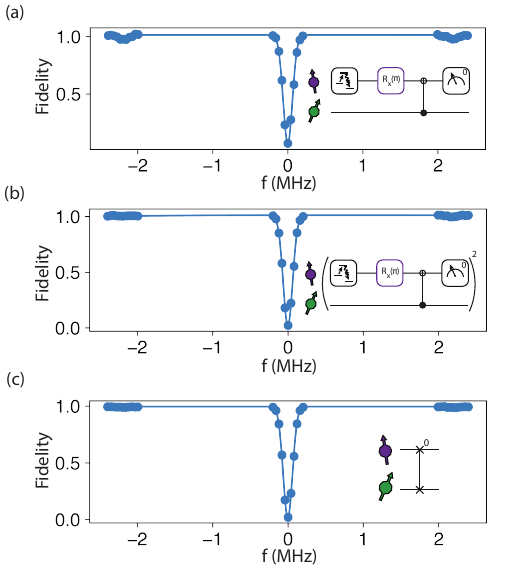}
\caption{\textbf{ESR spectra for different initialisation methods.} \textbf{a.} We use a single round of measurement based initialisation (MBI) to initialise the nitrogen spin. The data is fitted to the sum of three Gaussians: $a - A_1 \exp(-(x-x_1)^2/(2\sigma_1^2)) - A_2 \exp(-(x-x_2)^2/(2\sigma_2^2)) - A_3 \exp(-(x-x_3)^2/(2\sigma_3^2))$. We fit to Gaussian functions, because we used a Gaussian pulse shape to measure the ESR spectrum. We find the contrast as $A_2/(A_1+A_2+A_3)$ where $A_2$ is the amplitude of the middle dip. We find a value of 0.924(8) with a reduced chi-square of the fit of $\chi^2_r = 1.9$. \textbf{b.} We use two rounds of MBI to initialise the nitrogen spin. We find a contrast of 0.979(6) with $\chi^2_r = 3.7$. \textbf{c.} We used SWAP initialisation (Fig. \ref{figs3}) to initialise the nitrogen spin. We find an initialisation fidelity of 0.985(7) with $\chi^2_r = 21.6$.}
\label{figs2}
\end{figure*}

An alternative to SWAP is to use measurement-based initialisation (MBI). The sequence is given in the inset of Fig. \ref{figs2}a,b. First, we initialise the electron spin in $m_s = -1$, after which we apply a weak microwave pulse. This microwave flips the electron spin back to $m_s = 0$ only when the nitrogen spin is in $m_I = 0$. Finding the electron spin in $m_s = 0$ upon reading out, initializes the nitrogen spin in $m_I = 0$.

To compare the different initialisation methods, we measure an electron spin resonance (ESR) spectrum after initialisation of the nitrogen spin. In Fig. \ref{figs2}a, we perform a single round of MBI initialisation. In Fig. \ref{figs2}b, we perform two rounds of MBI intialisation. At the cost of a slower experimental rate, we find a significantly improved initialisation fidelity. In Fig. \ref{figs2}c, we show the ESR spectrum after SWAP initialisation as in Fig. \ref{figs3}. We find a marginally improved fidelity compared to double MBI initialisation.

\clearpage

\section{Nitrogen-spin readout} \label{nitrogen_readout}

\begin{figure*}[h]
\includegraphics[width=0.5\textwidth]{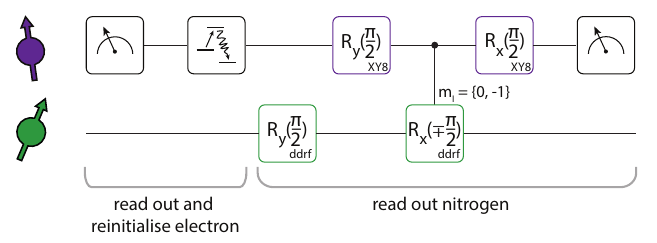}
\caption{\textbf{Sequence to read out the electron and nitrogen spin sequentially.} The electron spin is read out and initialised optically. The nitrogen spin is then mapped to the electron spin using only gates characterised with gate set tomography. Finally, we read out the electron spin.}
\label{figs4}
\end{figure*}

After each gate set tomography experiment performed in this work, we measure both the electron spin and the nitrogen spin. The full sequence to do so is given in Fig. \ref{figs4}. We read out the electron spin optically in a single shot \cite{Robledo2011}. Afterwards, we reinitialise the NV electron spin in $m_s = 0$ and read out the nitrogen spin. We compile our readout out of gates characterised by GST. The applied sequence of gates maps the nitrogen $z$-projection to the electron spin $z$-projection, after which we read out the electron spin optically. 

\clearpage

\section{Nitrogen-spin coherence} \label{nitrogen_coherence}

\begin{figure*}[h]
\includegraphics[width=0.7\textwidth]{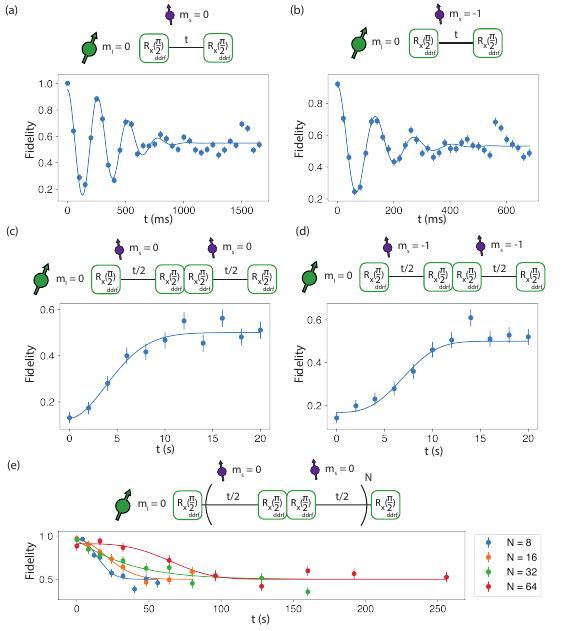}
\caption{\textbf{Nitrogen coherence.} \textbf{a.} Ramsey experiment with the electron in $m_s = 0$. We find $T_2^* = 553(46)$ ms ($\chi^2_r = 4.1$). \textbf{b.} Ramsey experiment with the electron in $m_s = -1$. We find $T_2^* = 172(24)$ ms ($\chi^2_r = 3.2$). \textbf{c.} Spin echo measurement with the electron in $m_s = 0$. We find $T_2 = 5.7(5)$ s ($\chi^2_r = 0.97$). \textbf{d.} Spin echo measurement with the electron in $m_s = -1$. We find $T_2 = 8.0(7)$ s ($\chi^2_r = 1.5$). \textbf{e.} Nitrogen coherence for different numbers of echo pulses $N$. For N = 8, we find $T_2 = 18(2)$ s ($\chi^2_r = 1.2$). For N = 16, we find $T_2 = 28(3)$ s ($\chi^2_r = 0.7$). For N = 32, we find $T_2 = 37(10)$ s ($\chi^2_r = 1.6$). For N = 64, we find $T_2 = 73(9)$ s ($\chi^2_r = 1.2$).}
\label{figs1}
\end{figure*}

In this section, we discuss coherence measurements of the nitrogen spin. To bring the nitrogen spin in a superposition and to apply echo pulses, we use the DDRF gates (Fig. \ref{figs1}). First, we measure the inhomogeneous dephasing time $T_2^*$ of the nitrogen spin for the electron spin in $m_s = 0$ (Fig. \ref{figs1}a) and $m_s = -1$ (Fig. \ref{figs1}b). We find that $T_2^*$ is dependent on the electron spin state, obtaining a factor $\sim 3$ larger $T_2^*$ in $m_s = 0$ compared to $m_s = -1$. We do not currently have a good explanation for this observed difference.

In Fig. \ref{figs1}c we show a nitrogen spin echo measurement with the electron spin in $m_s = 0$. In Fig. \ref{figs1}d we show the same measurement with the electron spin in $m_s = -1$. Here, we find that the coherence time in $m_s = -1$ is longer than in $m_s = 0$, which is the opposite finding from the $T_2^*$ measurement presented above. The observed increase in $T_2$ time for $m_s = -1$ may have to do with the presence of a frozen core around the NV center, which implies that more quasi-static noise can be echoed out. 

In Fig. \ref{figs1}e, we show the nitrogen coherence for different numbers of echo pulses. We find an increase of coherence with increasing number of pulses, as expected. For $N=64$, we find a coherence time of $T_2 = 73(9)$ s.

\clearpage

\section{Electron-spin coherence} \label{electron_coherence}

\begin{figure*}[h]
\includegraphics[width=\textwidth]{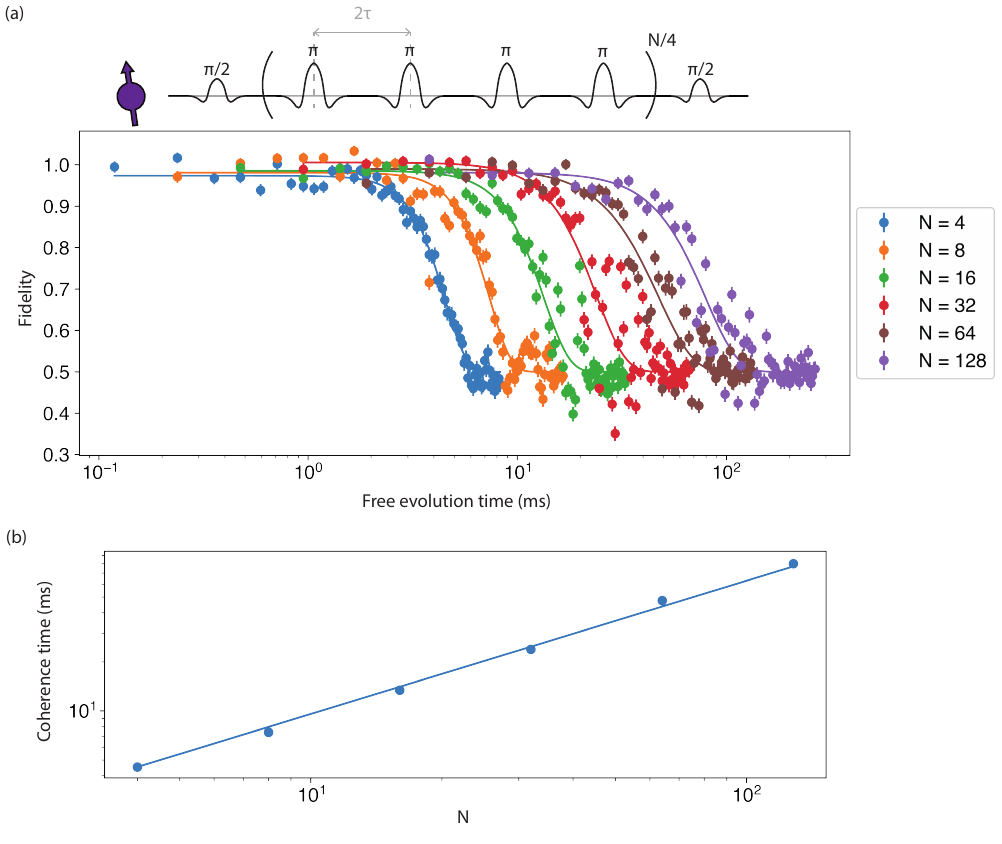}
\caption{\textbf{Electron coherence.} \textbf{a.} (Top) Sequence to measure the electron coherence. We prepare the electron in a superposition and apply a number of $\pi$-pulses $N$. (Bottom) Experimental result. From $N=4$ to $N=128$ we find respectively $T_2 = 4.53(4)$ ms ($\chi^2_r = 1.9$), $T_2 = 7.4(2)$ ms ($\chi^2_r = 7.5$), $T_2 = 13.4(4)$ ms ($\chi^2_r = 9.5$), $T_2 = 23.9(9)$ ms ($\chi^2_r = 17.7$), $T_2 = 48(2)$ ms ($\chi^2_r = 9.3$), $T_2 = 80(3)$ ms ($\chi^2_r = 7.7$). \textbf{b.} Scaling of the coherence time with the number of pulses $N$. We fit to $T_{N=4} (N/4)^\eta$ where $T_{N=4}$ is the coherence time for $N=4$. We find $\eta=0.82(2)$ ($\chi^2_r = 27.3$).}
\label{figs8}
\end{figure*}

In this section, we discuss the electron coherence under dynamical decoupling. In Fig. \ref{figs8}a, we show the electron coherence as a function of the number of $\pi$-pulses applied to the electron spin. We make sure that the interpulse delay $\tau$ is a multiple of the period set by the $^{13}$C Larmor frequency to mitigate resonances from electron-nuclear interaction. While we see a steady increase in coherence time for increasing number of pulses (Fig. \ref{figs8}b), significant outliers in the data are also visible. We attribute these to the presence of $50$ Hz noise. While detrimental at large values of $\tau$, at $\tau=7.304$ \textmugreek s, which is the value of $\tau$ used for the gate designs in this work, we are not significantly affected by this.

\clearpage

\section{Gate parameter calibrations} \label{gate_calibrations}

In this section, we discuss the calibrations performed for each gate. Next to magnetic field adjustment (Section \ref{B_field}) and balancing of the I \& Q channels of the IQ modulator, we perform amplitude calibration of each single-qubit gate. The amplitude of the RF pulses for the two-qubit gate is set equal to those of the single-qubit gates and is not separately calibrated. The only difference constitutes a $\pi$-phase shift of half of the RF pulses (Fig. 2).

\begin{figure*}[h]
\includegraphics[width=\textwidth]{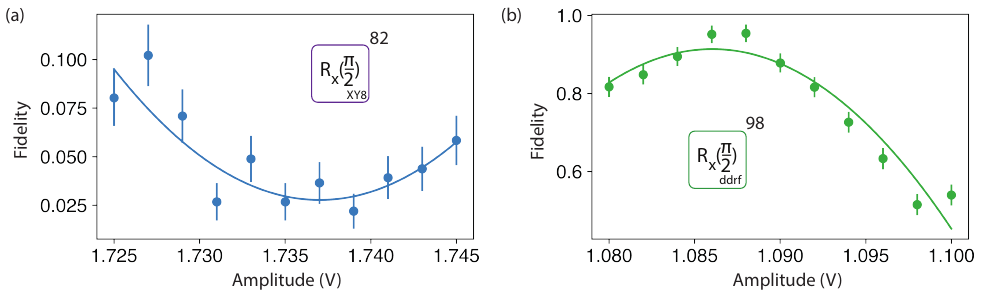}
\caption{\textbf{Amplitude calibration of electron and nitrogen gates.} \textbf{a.} We apply $82$ electron $\pi/2$ gates while sweeping the amplitude of the $\pi/2$ pulse. By fitting to $A + a(x-x_0)^2$, we obtain the optimal amplitude of $1.7370(9)$ V ($\chi^2_r = 1.4$). \textbf{b.} We apply $98$ nitrogen $\pi/2$ gates while sweeping the RF amplitude in the DDRF gate. We find the optimal amplitude at $1.0860(9)$ V ($\chi^2_r = 3.6$).}
\label{figs9}
\end{figure*}

In Fig. \ref{figs9}, we show two examples of amplitude calibration. In Fig. \ref{figs9}a we apply $82$ electron $\pi/2$ pulses and read out the electron spin while the nitrogen spin is in $m_I = 0$. We vary the amplitude of the $\pi/2$ pulse in the gate. The minimum is found by fitting a parabola to obtain the optimal amplitude. The $\pi$ pulses that make up the XY8 sequence are calibrated similarly. In Fig. \ref{figs9}b we apply $98$ nitrogen $\pi/2$ pulses with DDRF and read out the nitrogen spin. We vary the amplitude of the RF driving of the nitrogen spin. The maximum value of the fit gives the optimal amplitude for the RF driving. 

\clearpage

\section{Single-shot readout correction}
\label{sec:SSRO}
For the results shown in Figures 2 and 4 of the main text as well as in many supplementary figures (Figs. \ref{figs11}, \ref{figs2}, \ref{figs1}, \ref{figs8}, \ref{figs9}, \ref{fig:figs3_rb}, \ref{fig:figs2_rb}, \ref{fig:figs4_rb}, \ref{fig:figs6_swap}, \ref{fig:figs15}, \ref{fig:figs14}), we correct the electron measurement results for known readout errors. We do this by using single-shot readout (SSRO) correction following Pompili et al. \cite{Pompili2021}. Before and during our measurements, we run SSRO calibrations to find $F_0$ = $P(\text{measure } m_s = 0 | \text{state is } m_s = 0)$ and similarly $F_1$. Now the measurement of the electron spin can be described by

\begin{equation} \label{eq:ssro_cor}
\textbf{m} = 
\begin{pmatrix}
F_0 & 1-F_1 \\
1-F_0 & F_1

\end{pmatrix}
\textbf{p}
\end{equation}

Here, \textbf{m} $ = \begin{pmatrix} m_0 & m_1 \end{pmatrix}^{\textrm{T}}$ is a vector with measured populations and \textbf{p} $= \begin{pmatrix} p_0 & p_1 \end{pmatrix}^{\textrm{T}}$ is a vector with the true populations we expect based on the measurement fidelities. Measuring $m_0$, $m_1$, $F_0$ and $F_1$, we obtain $p_0$ and $p_1$ by matrix inversion. For example, for 500 repetitions of preparing the electron spin in $m_s$ = 0 and reading it out, one could measure 406 occurrences of $m_s = 0$. This gives $m_0 = 406/500$ and $m_1 = 94/500$. Using typical values of $F_0 = 82\%$ and $F_1 = 99\%$ this gives $p_0 = 0.99$ and $p_1 = 1 - p_0 = 0.01$. Note that this method can return non-physical central values due to noise (e.g. a population number exceeding one). An alternative method to do SSRO correction is Iterative Bayesian Unfolding \cite{Nachman2020}.

For readout of the nitrogen spin, its state is mapped to the electron spin and subsequently measured as an electron spin state (see Supplementary Note \ref{nitrogen_readout}). Therefore we also use the correction described in this section for measurements of the nitrogen spin state. Note that we do not correct for the infidelity of the readout circuit of the nitrogen spin, which explains why the fidelities are typically not perfect (see e.g. Fig. \ref{figs1}).

The error bars on the data in the relevant figures (Figs. 2, 4, \ref{figs11}, \ref{figs2}, \ref{figs1}, \ref{figs8}, \ref{figs9}, \ref{fig:figs3_rb}, \ref{fig:figs2_rb}, \ref{fig:figs4_rb}, \ref{fig:figs6_swap}, \ref{fig:figs15}, \ref{fig:figs14}) represent one standard deviation and are calculated using the measured populations $m_0$, $m_1$. The uncertainty on $m_0$, $m_1$ is binomial:

\begin{equation}
    \sigma_{m_0} = \sigma_{m_1} = \sqrt{\frac{m_0(1-m_0)}{N}},
\end{equation}

where $N$ is the number of experimental repetitions. We can invert equation \ref{eq:ssro_cor} to obtain the error on the expected populations \cite{Pompili2021}:

\begin{equation}
    \sigma_{p_0} = \sigma_{p_1} = \frac{\sigma_{m_0}}{F_0+F_1-1}.
\end{equation}

\clearpage

\section{Randomized benchmarking}
\label{sec:RB}

\begin{figure*}[b]
    \includegraphics[width=\textwidth]{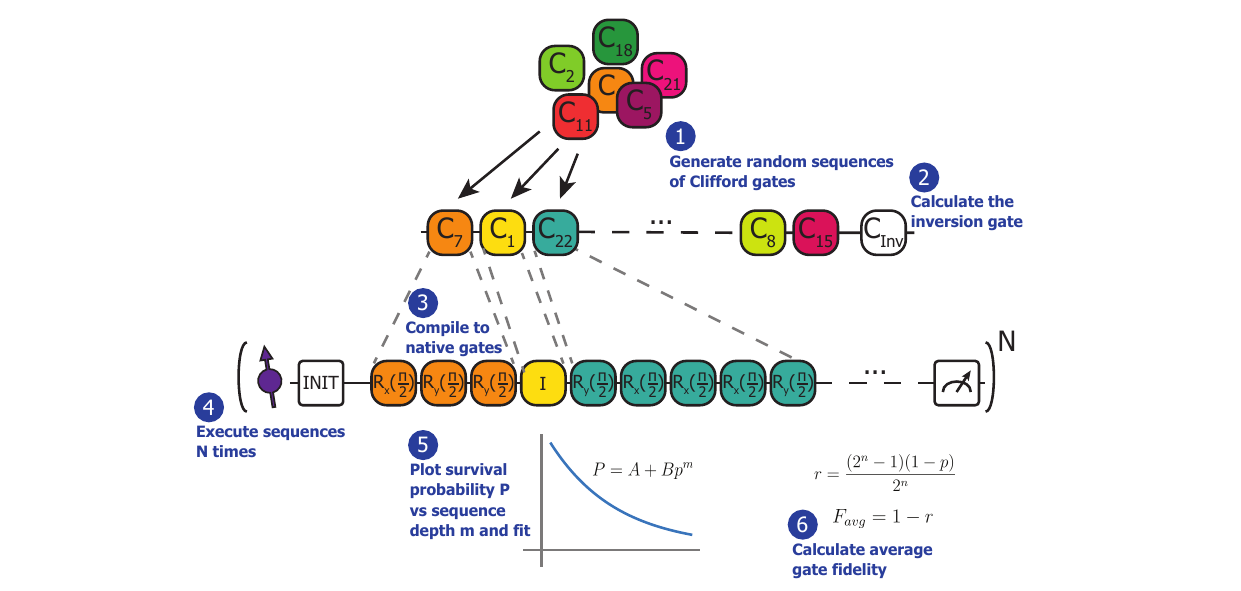}
    \caption{\textbf{Illustration of the Clifford randomized benchmarking protocol.}}
    \label{fig:RB_protocol}
\end{figure*}

A common method to characterize quantum gates is randomized benchmarking (RB) \cite{Emerson2005,Magesan2011,Helsen2022,Wallman2014,Proctor2017,Epstein2014}. Randomized benchmarking gives a metric for how well a quantum state `survives' sequences of random quantum gates. In contrast to the GST that we use in the main text, randomized benchmarking does not provide the full process matrix. Here, we perform (single-qubit) Clifford RB and compare against the results obtained from single-qubit GST. The protocol is as follows (see also the illustration in figure \ref{fig:RB_protocol}):

\begin{enumerate}
    \item Random sequences of Clifford gates of different depths (lengths) are generated.
    \item An inversion gate (also a Clifford gate) is calculated and appended to the random sequence. For half of the sequences, the inversion gate theoretically makes the total sequence equal to identity. For the other half, the inversion gate additionally incorporates a $\pi$ pulse, rotating the measurement basis.
    \item The Clifford gates are compiled out of native gates (Identity, $X(\pi/2)$, $Y(\pi/2)$). The native gates were chosen to be the same as the gates characterized with GST. The average number of native gates per Clifford gate is $N = 3.125$.
    \item The compiled sequences are run on either the initialized electron spin (500 repetitions) or the initialized nuclear spin (1000 repetitions). The obtained counts are corrected for electron readout fidelity (Section \ref{sec:SSRO}).
    \item We plot the survival probability $P$ of the quantum state as a function of native gate sequence depth $m$. We fit $P = A + Bp^m$ to the average survival probabilities per depth, taking the binomial uncertainties on the points (see Section \ref{sec:SSRO}) as relative weights. We extract the depolarizing parameter $p$. $A$ and $B$ are values that capture the state preparation and measurement (SPAM) errors. $A$ is fixed to 0.5.
    \item The average gate infidelity $r$ is calculated using $r = \frac{(2^n-1)(1-p)}{2^n}$ where $n$ is the number of qubits. The average gate fidelity is $F_{avg} = 1 - r$.
    
\end{enumerate}

To be able to compare the results from RB and GST, we use a simulation to generate RB data based on the process matrices of the native gates, that were obtained using GST \cite{Blume-Kohout2017}. The simulated RB data is analysed in a similar fashion as the experimental RB data. Then, the average gate fidelities of both methods can be compared. GST gives gate fidelities for the native gates separately. As a sanity check, we took the average of these gate fidelities, weighted by their occurrence in the RB sequences used for both experiment and simulation, and compared the weighted average to the RB simulation. In all three cases described below, this weighted average was - within error - the same as the result of the RB simulation. This indicates that one can directly compare the weighted average gate fidelity of the GST report to an average gate fidelity found with RB.

\subsection{RB on the electron spin}
We generate 30 random sequences of Clifford gates for each depth $m_C \in$ \{5, 10, 20, 50, 100, 200, 500, 750\} Clifford gates. The inversion gate (also a Clifford gate) theoretically returns the quantum state to the measurement basis, which we alternate between the electron state $m_s = 0$ and $m_s = -1$. 

\begin{figure*}
    \includegraphics[width=0.8\textwidth]{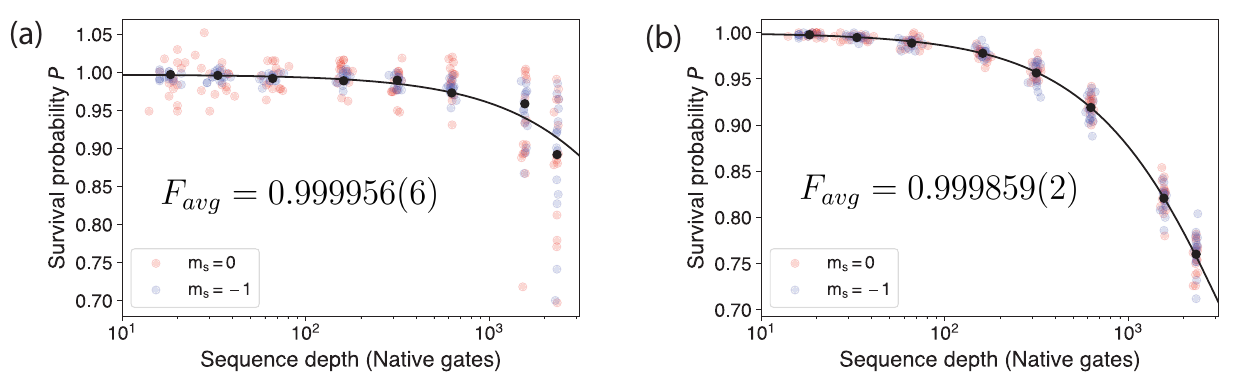}
    \caption{\textbf{Electron RB results.} The nitrogen spin is initialized in $m_I=0$ and no decoupling sequences are used around the $\pi/2$ gates. Red (blue) dots are values resulting from a sequence ending with an inversion gate bringing the spin to $m_s=0$ ($m_s=-1$). Black dots are the average of all survival probabilities belonging to one Clifford depth $m_C \in \{5, 10, 20, 50, 100, 200, 500, 750\}$. The black line is a fit to the black dots. Errorbars on the black dots are binomial errors and smaller than the datapoints. \textbf{a.} Experimental result. The depolarizing parameter is $p = 0.99991(1)$ ($\chi^2_{r} = 40.7$), resulting in an average gate fidelity of $F_{avg} = 0.999956(6)$. \textbf{b.} Simulation of the RB experiment based on the process matrices obtained with GST. The depolarizing parameter $p = 0.999717(4)$ ($\chi^2_{r} = 1.30$) results in an average gate fidelity of $F_{avg} = 0.999859(2)$. }
    \label{fig:figs3_rb}
\end{figure*}

Figure \ref{fig:figs3_rb}a shows the survival probability for each sequence. For this experiment, the nitrogen spin was initialised in $m_I=0$. The identity gate consists of an XY8 decoupling sequence (Fig. 2a). The $\pi/2$ gates around $x$ and $y$ are Hermite pulses without XY8 decoupling sequences around them (Fig. 3a). We find a depolarizing parameter $p = 0.99991(1)$ ($\chi^2_{r} = 40.7$). The average gate fidelity of a native gate is $F_{avg} = 0.999956(6)$.

With GST, we also investigated this experimental regime with the nitrogen initialized in $m_I=0$ and no decoupling sequences around the $\pi/2$ gates. The results are shown on the right hand side of Fig. 3b. With the process matrices obtained from this GST characterization, we simulate RB data (with a binomial spread) using the pygsti package \cite{pygsti}. For this, the same random sequences are used as for the actual RB experiment. The results are shown in Fig. \ref{fig:figs3_rb}b. We found a depolarizing parameter $p = 0.999717(4)$ ($\chi^2_{r} = 1.30$), resulting in an average gate fidelity of $F_{avg} = 0.999859(2)$. This is in correspondence with the weighted (for their occurrence in the RB sequences) average of the gate fidelities from GST, which is 0.99985(3).

Interestingly, the average gate fidelity found with experimental RB ($F_{avg} = 0.999956(6)$) is significantly higher than the average gate fidelity reported by the experimental GST ($F = 0.99985(3)$). This can be an indication of non-Markovianity in our system, for example in the form of slow fluctuations of the magnetic field. This type of error manifests as a coherent error that is the same within one measurement repetition, but differs from repetition to repetition. Since GST amplifies coherent errors, it is relatively sensitive to such low-frequency noise, whereas the random nature of RB sequences makes it relatively insensitive \cite{Blume-Kohout2017}. The large spread in survival probabilities for the actual RB experiment (especially for long sequences) may also be related to the non-Markovianity in our system. The large $\chi^2_{r}$ can also be an indication of the noise being non-Markovian \cite{Wallman2018,Helsen2022}. 

\begin{figure*}
    \includegraphics[width=0.8\textwidth]{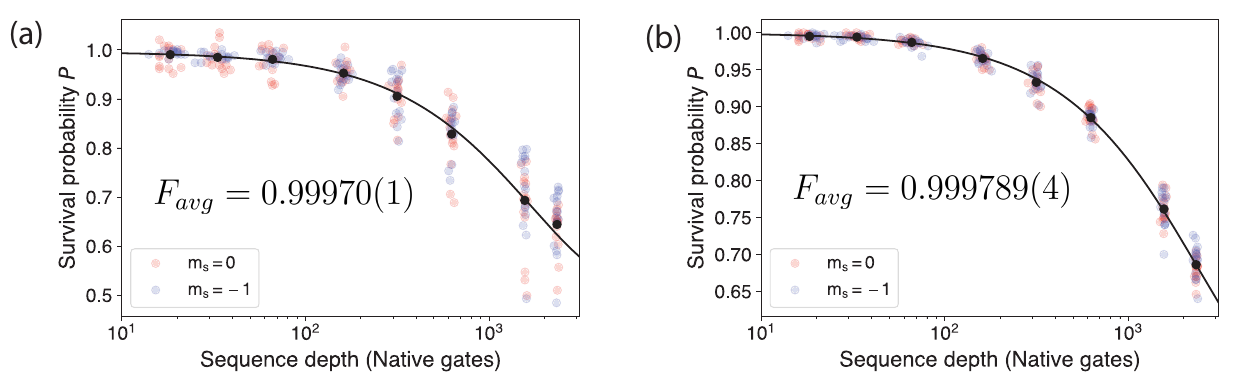}
    \caption{\textbf{Electron RB results.} The nitrogen spin is initialized in $m_I=0$ and there are XY8 decoupling sequences around the $\pi/2$ gates. Red (blue) dots are values resulting from a sequence ending with an inversion gate bringing the spin to $m_s=0$ ($m_s=-1$). Black dots are the average of all survival probabilities belonging to one Clifford depth $m_C \in \{5, 10, 20, 50, 100, 200, 500, 750\}$. Black line is a fit to the black dots. Errorbars on the black dots are binomial errors and smaller than the datapoints. \textbf{a.} Experimental results. The depolarizing parameter is $p = 0.99941(2)$ ($\chi^2_{r} = 11.6$), resulting in an average gate fidelity of $F_{avg} = 0.99970(1)$. \textbf{b.} Simulation of RB experiment based on the process matrices obtained with GST. The depolarizing parameter $p = 0.999578(8)$ ($\chi^2_{r} = 2.42$) results in an average gate fidelity of $F_{avg} = 0.999789(4)$.}
    \label{fig:figs2_rb}
\end{figure*}

The results in Fig. \ref{fig:figs2_rb} are from a similar experiment. Except here, there are XY8 sequences around the $\pi/2$ pulses around $x$ and $y$ (Fig. 2a). The nitrogen was initialised in $m_I=0$. From the fit, we obtain a depolarizing parameter $p = 0.99941(2)$ ($\chi^2_{r} = 11.6$). The average gate fidelity of a native gate is $F_{avg} = 0.99970(1)$. Figure \ref{fig:figs2_rb}b shows simulated data based on the process matrices of the gates, that were obtained by the GST experiment reported in Fig. \ref{figs5}b, second column. We find a depolarizing parameter $p = 0.999578(8)$ ($\chi^2_{r} = 2.42$). The average gate fidelity of a native gate is $F_{avg} = 0.999789(4)$. This value is the same as the weighted average of the gate fidelities of the GST report, which is 0.99979(5).

Again, there is a discrepancy between the average gate fidelity found with RB and the average gate fidelity found with GST. However, it is significantly smaller, possibly because the XY8 decoupling sequences around the Hermite pulses reduce the non-Markovianity in our system.

\begin{figure*}
    \includegraphics[width=0.8\textwidth]{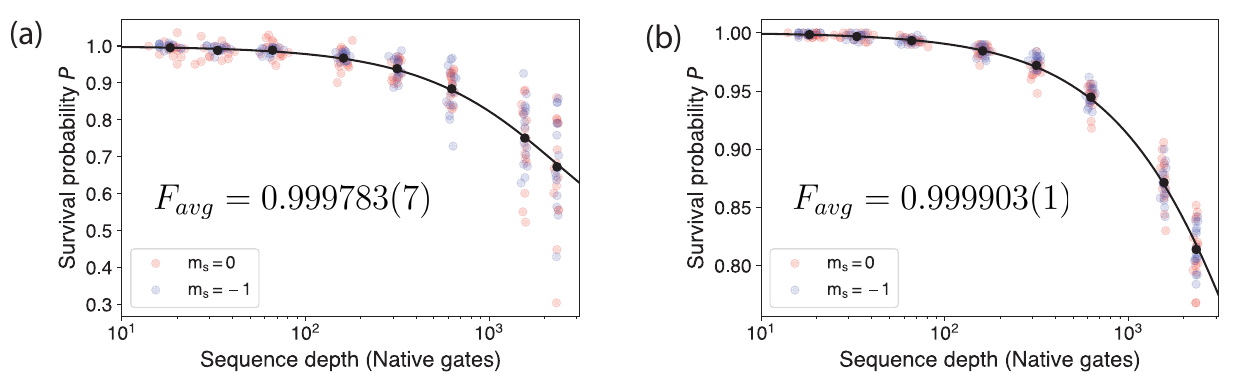}
    \caption{\textbf{Electron RB results.} The nitrogen spin is initialized in $m_I=0$ and there are XY8 decoupling sequences around the $\pi/2$ gates, with $\tau = 2.0 $\textmugreek s. Red (blue) dots are values resulting from a sequence ending with an inversion gate bringing the spin to $m_s=0$ ($m_s=-1$). Black dots are the average of all survival probabilities belonging to one Clifford depth $m_C \in \{5, 10, 20, 50, 100, 200, 500, 750\}$. Black line is a fit to the black dots. Errorbars on the black dots are binomial errors and smaller than the datapoints. \textbf{a.} Experimental results. The depolarizing parameter is $p = 0.99957(1)$ ($\chi^2_{r} = 6.56$), resulting in an average gate fidelity of $F_{avg} = 0.999783(7)$. \textbf{b.} Simulation of RB experiment based on the process matrices obtained with GST. The depolarizing parameter $p = 0.999807(3)$ ($\chi^2_{r} = 0.81$) results in an average gate fidelity of $F_{avg} = 0.999903(1)$.}
    \label{fig:rb_fig_tau2us}
\end{figure*}

Fig. \ref{fig:rb_fig_tau2us}a shows the results of another RB experiment with XY8 sequences around the $\pi/2$ pulses. Here, we use a decoupling time $\tau$ of 2.0 \textmugreek s. We find a depolarizing parameter $p = 0.99957(1)$ ($\chi^2_{r} = 6.56$), resulting in an average gate fidelity of $F_{avg} = 0.999783(7)$. Figure \ref{fig:rb_fig_tau2us}b shows simulated data based on the process matrices of the gates, that were obtained by doing the GST experiment of which the results are shown in Fig. \ref{figs5}b, third column. The simulation gives a depolarizing parameter $p = 0.999807(3)$ ($\chi^2_{r} = 0.81$) and an average gate fidelity of $F_{avg} = 0.999903(1)$. This value is the same as the weighted average of the gate fidelities of the GST report, which is 0.99990(4).

\subsection{RB on the nitrogen spin}
\begin{figure*}
    \includegraphics[width=0.8\textwidth]{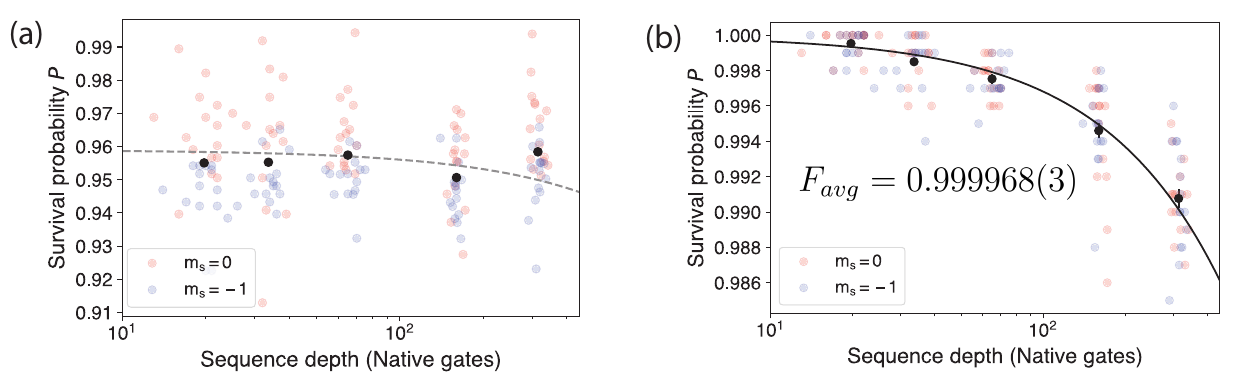}
    \caption{\textbf{Nitrogen RB results.} The electron was initialized in $m_s=0$ and DDRF gates were used (Fig. 2a). Red (blue) dots are values resulting from a sequence ending with an inversion gate bringing the nitrogen spin to the state that is mapped to $m_s=0$ ($m_s=-1$) during readout. Black dots are the average of all survival probabilities belonging to one Clifford depth $m_C \in \{5, 10, 20, 50, 100\}$. \textbf{a.} Experimental result. The RB experiment took $\sim 3$ hours compared to $\sim 1.5$ hours for the GST results in Fig. 3c, right hand side. There is no decay that could be fit. The dotted line is an exponential decay with the decay parameter obtained from the simulation in panel b, for comparison. \textbf{b.} Simulation of RB experiment based on the process matrices obtained with GST. Black line is a fit to the black dots, yielding a depolarizing parameter of $p = 0.999937(6)$ ($\chi^2_{r} = 3.12$). This gives an average gate fidelity of $F_{avg} = 0.999968(3)$.}
    \label{fig:figs4_rb}
\end{figure*}

The results for RB on the nitrogen spin are shown in Fig. \ref{fig:figs4_rb}a. Here, the electron is in an eigenstate ($m_s=0$) and we use DDRF gates (Fig. 2a). As expected, there is no visible decay for these depths. This data should be compared to the GST results shown on the right hand side of Fig. 3c. Using the process matrices obtained with that GST experiment, we simulate the RB experiment in Fig. \ref{fig:figs4_rb}b. We find $p = 0.999937(6)$ ($\chi^2_{r} = 3.12$). The average gate fidelity of a native gate is $0.999968(3)$. The weighted average of the gate fidelities from the GST report is 0.99996(3).

\clearpage

\section{SWAP: comparison to the GST prediction}
In Figure 4b of the main text, we apply the compiled SWAP gate $n$ times. After an even number of SWAPs, we then measure the average state fidelity of the six cardinal states:
\begin{equation}
    F_{\mathrm{avg}} = \frac{1}{6}\sum_{i=1}^6 \bra{\psi}_i \rho_{i} \ket{\psi}_i,
\end{equation}
with $\ket{\psi}_i \in \{\ket{Z},\ket{-Z},\ket{X},\ket{-X},\ket{Y},\ket{-Y}\}$, and with $\rho_i$ the experimentally obtained state (ideally $\rho_i = \ket{\psi}_i\bra{\psi}_i$).

For this purpose, the electron is initialized to each of the cardinal axes and the nitrogen is initialised in $m_I=0$. At the end of an even number of SWAP gates, we measure the projections of the electron spin onto all cardinal axes, and the nitrogen spin along $z$.  

\begin{figure}[h]
    \includegraphics[width = \textwidth]{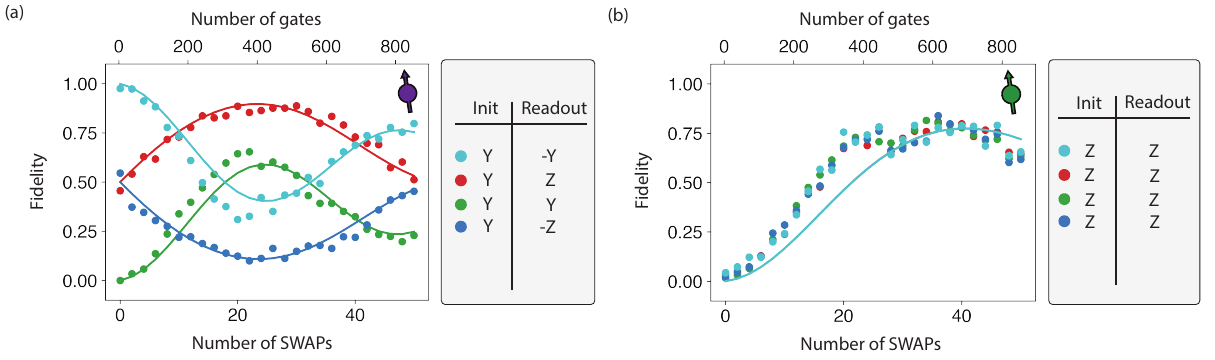}
    \caption{\textbf{Comparison of GST-based prediction to experiment for different spin components.} \textbf{a.} Electron spin population measured along different cardinal axes (points) and corresponding GST predictions (lines). \textbf{b.} Nitrogen spin population measured along \textit{z} (points) and corresponding GST predictions (lines). The error bars are smaller than the data points.}
    \label{fig:figs6_swap}
\end{figure}

The average state fidelity after $n$ SWAP gates is well reproduced by the process matrices obtained from two-qubit GST. Moreover, the action of the full gate sequence on the individual spin components of both electron and nitrogen spin can be reproduced. This is illustrated in Figure \ref{fig:figs6_swap} which shows an example of state preparation and measurement in different bases and its comparison to the GST prediction. The results corroborate the accuracy of the process matrices obtained from GST.

Even though the average state fidelity after $n=20$ SWAP operations reaches 0.5, the individual electron spin components clearly show coherent rotations well beyond that, which are expected to come from coherent errors. Together with the predictive power of GST, this can be used to design tailored error mitigation to cancel the effect of accumulated coherent errors in specific extended gate sequences (Section \ref{swap_error_mitigation}). 

\clearpage
\section{SWAP: error mitigation techniques}\label{swap_error_mitigation}

As illustrated in Figure 4b of the main text, the average fidelity of the six cardinal electron states after an even number of SWAPs is mainly limited by a build-up of coherent errors. The particular sequence is close to a worst-case scenario, as the same gate is repeated $n$ times, allowing coherent errors to add up with circuit depth. As circuits of practical interest are in general not random, the effect of coherent errors on the results will likely be significant. The direct solution is to improve the precision of the calibration of the basic gate parameters in order to further reduce the coherent errors. However, this can be challenging at the high-fidelity levels achieved here.

\begin{figure}[h]
    \includegraphics[width = 0.8\textwidth]{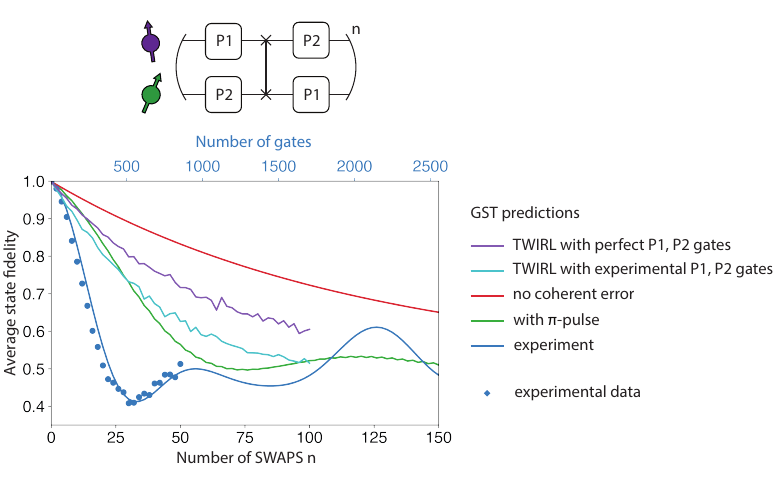}
    \caption{\textbf{Theoretical analysis of error mitigation techniques on the electron-nitrogen SWAP gate sequence.} Similar to Figure 4b of the main text, with the addition of Pauli twirling to mitigate the build-up of coherent errors. Top: Pauli Twirling compilation for the SWAP gate. For each realisation and each SWAP we choose $P_1$ and $P_2$ at random. For perfect Pauli operations, this block is identical to a simple SWAP. Bottom: Comparison of different mitigation techniques for the $n$ sequential SWAP gate sequence. Both electron $\pi$-pulse echo (green) and Pauli twirling with experimental gates for $P_1$, $P_2$ (light blue, experimental gates as characterised in Fig. 2) or perfect gates for $P_1$, $P_2$ (purple) are simulated to improve the average state fidelity. The error bars are smaller than the data points.}
    \label{fig:figs15}
\end{figure}

We highlight two different solutions to this issue using GST process matrix simulations, the results of which are illustrated in Figure \ref{fig:figs15}. First, using GST, we can determine the accumulated coherent error of a particular larger circuit block. In principle, this could allow one to tailor a specific gate sequence that cancels the effect of the coherent error completely. In the particular sequence at hand, a significant portion of the coherent error is due to a $Z_e I_n$ error during the controlled two-qubit gate. Therefore, echo-pulses on the electron after each swap (compiled as two electron $X_{\pi/2}$ gates), can significantly decrease the accumulated coherent error (Fig. \ref{fig:figs15}).

Second, we can use error mitigation techniques such as Pauli-twirling which was recently also used in the context of error extrapolation \cite{Li2017,Temme2017,Endo2018,Kim2023}. Due to the twirling, the coherent errors cannot add up linearly. In our case, we can build a Pauli-twirling set from our characterized gates as $\{I,X_{\pi/2}*X_{\pi/2},Y_{\pi/2}*Y_{\pi/2},X_{\pi/2}*Y_{\pi/2}*X_{\pi/2}*Y_{\pi/2}\}$. The twirling operations are compiled as illustrated in Figure \ref{fig:figs15}. For the simulation, we average over ten different realisations, picking a Pauli $P_1$ and $P_2$ randomly from the twirling set for each SWAP in the sequence. As we can see in the figure, the Pauli-twirling does not allow for a coherent build-up of errors, resulting in an exponential decay of the average fidelity of the six cardinal states. Comparing to ideal Pauli gates, we note that the effectiveness of this error mitigation technique relies on high-fidelity single qubit gates. These theoretical simulations for the example case (repeated SWAPs) illustrate that the GST characterisation can be used to determine tailored error mitigation methods to improve the fidelity of circuit blocks. 

\clearpage
\section{Quantum memory: electron-spin depolarisation during nitrogen-spin decoupling}

\begin{figure}[h]
    \includegraphics[width = 0.7\textwidth]{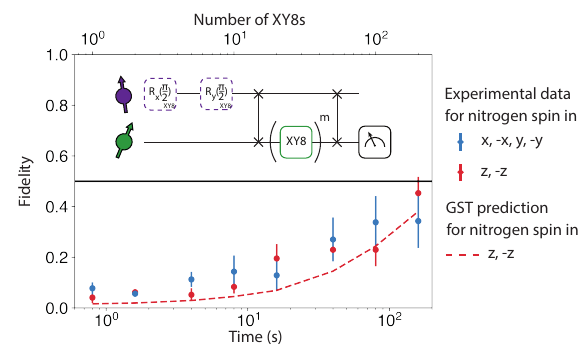}
    \caption{\textbf{Electron spin population during nitrogen decoupling.} After storing a quantum state ($\pm x$, $\pm y$ or $\pm z$) on the nitrogen spin and swapping it back to the electron spin, we measure the nitrogen spin to determine what happened to the electron spin during the decoupling sequence on the nitrogen spin. The electron spin population decays for increasing numbers of XY8s. The timescale is similar to the decoherence of the nitrogen spin in Fig. 4c, suggesting that control errors in the DDRF $\pi/2$ gates used to construct the nitrogen spin XY8 decoupling sequence lead to the observed decay.}
    \label{fig:figs14}
\end{figure}

In this section, we discuss additional data accompanying Fig. 4c. In Figure 4c, we read out the electron spin, which gives information about the state that was stored on the nitrogen spin during the main evolution time. In Figure \ref{fig:figs14}, we read out the nitrogen spin, which gives information about what happened to the electron spin during the decoupling sequence on the nitrogen spin. We find that the electron spin population decays during XY8 decoupling of the nitrogen spin. A similar decay is predicted from the two-qubit GST process matrices, even though we did not include electron-spin dephasing during the free evolution time in this prediction, as this depends on the detailed, microscopic spin-bath dynamics. The electron spin depolarisation occurs at the same timescale as the nitrogen spin decoherence in Fig. 4c, which suggests that electron-spin control errors in the DDRF gate are the underlying cause of the observed nitrogen spin decoherence, instead of direct decoherence of the nitrogen spin due to the surrounding spin bath.

\clearpage

\section{Additional single-qubit GST results}\label{additional_gst_results}

In this section, we show a range of additional gate-set-tomography results that complement those in the main text. In the electron-nitrogen system of the NV center, there are many possible modes of operation depending on how each qubit is used at a particular point in an algorithm (e.g. nitrogen spin initialised or mixed), for which different gate implementations can be used (e.g. including or excluding XY8 decoupling sequences). Therefore there is a variety of situations that are useful to characterise.

\subsection{Single-qubit GST on the electron spin}

\begin{figure*}[h]
\includegraphics[width=\textwidth]{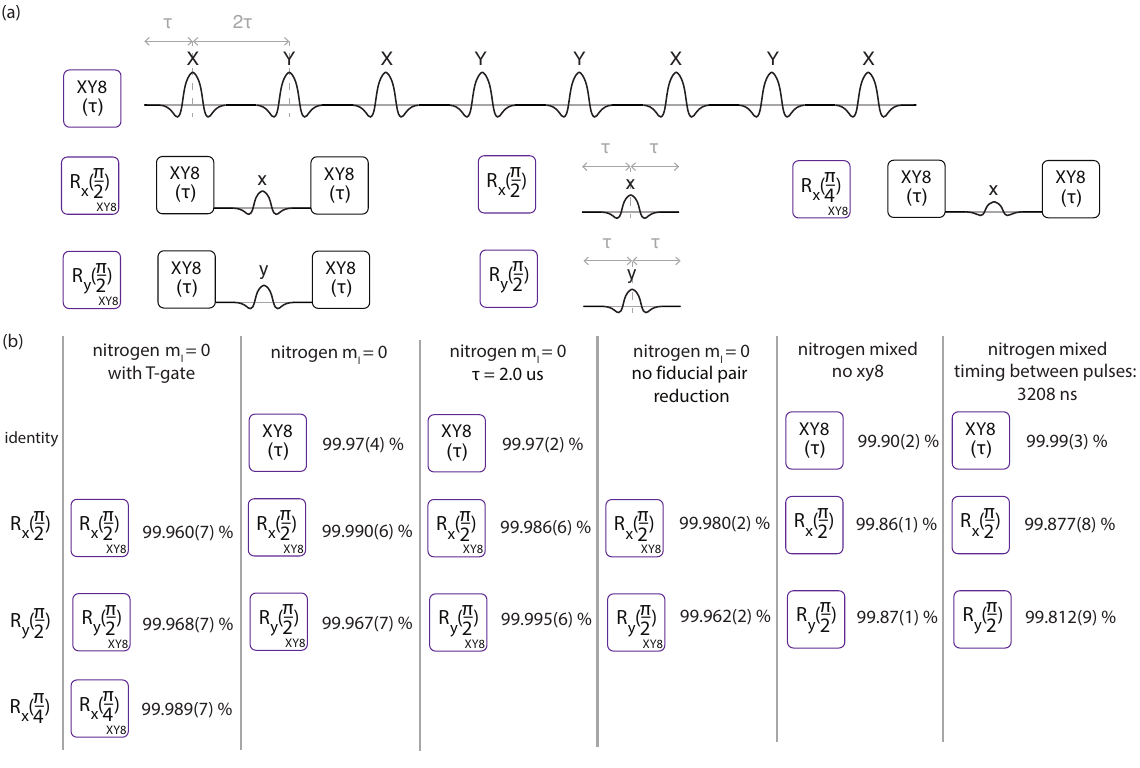}
\caption{\textbf{Additional electron single-qubit GST results.} \textbf{a.} The collection of electron gates applied in the various GST experiments. The XY8 sequence consists of multiple Hermite $\pi$ pulses with an interpulse delay of $\tau = 7.304$ \textmugreek s unless indicated otherwise. Next to the gates in the main text, we also calibrate an electron $\pi/4$ gate around $x$. \textbf{b.} Average gate fidelities for the additional electron single-qubit GST results. (column 1) GST on the electron gates with an additional T-gate or $\pi/4$ gate. (column 2) Electron gates with XY8 with the nitrogen spin in $m_I = 0$. (column 3) Electron gates with XY8 for an interpulse delay of $\tau = 2.0$ \textmugreek s instead of $\tau = 7.304$ \textmugreek s. (column 4) Electron gates with XY8 without fiducial pair reduction. (column 5) Electron gates without XY8 with the nitrogen spin in a mixed state. The time delay between two $\pi/2$ pulses is 1344 ns. (column 6) Electron gates without XY8 with the nitrogen in a mixed state. We apply a specific timing (3208 ns) between two consequent $\pi/2$ pulses, which is a multiple of the electron-nitrogen interaction time.}
\label{figs5}
\end{figure*}

In Fig. \ref{figs5}, we show six additional single-qubit GST results for the electron spin. Fig. \ref{figs5}a shows how each electron gate is compiled out of electron pulses. In Fig. \ref{figs5}b, we show the average gate fidelities of each gate for different scenarios. While we only show and discuss the obtained average gate fidelities, we obtained a full process matrix of each gate. These are provided in the published data accompanying this paper. 

The first column in Fig. \ref{figs5}b shows the single-qubit GST results for the T-gate or $\pi/4$ gate, which we include to explicitly show a universal gate set. The nitrogen-spin state is $m_I = 0$ and the implemented T-gate is based on the $\pi/2$ gate, but has approximately half the amplitude. The T-gate fidelity is similar to the electron $\pi/2$ gates.

In the second column, we show GST results for the electron gates with XY8 deoupling and with the nitrogen initialised in $m_I = 0$. Compared to the mixed electron-spin state in Fig. 3b, the fidelities are significantly improved for both $\pi/2$ gates. Compared to the simple direct microwave pulse (main text Fig. 3b), the fidelity is slightly lower, while this gate is significantly longer (see main text). Note that for this situation and gate we also performed a randomized benchmarking experiment (as shown in Fig. \ref{fig:figs2_rb}).

The value $\tau = 7.304$ \textmugreek s was chosen in the gate design to avoid unwanted nitrogen-spin rotations (Section \ref{B_field}). However, when only controlling the electron spin this requirement can be relaxed. The third column shows a GST result for a shorter $\tau = 2.0$ \textmugreek s, which may be expected to reduce dephasing due to the shorter gate length. However, we find comparable fidelities to $\tau = 7.304$ \textmugreek s (second column), suggesting that decoupling at a different $\tau$ has no large influence on the fidelity of the electron-spin gates in the single-qubit space.

In the fourth column in Fig. \ref{figs5}b, we perform the same experiment as in the second column, but without fiducial pair reduction and without characterising the identity gate. Fiducial pair reduction is a method to reduce the number of quantum circuits that have to be measured, making a gate set tomography experiment significantly more efficient \cite{Nielsen2021,Rudinger2023}. Unless mentioned otherwise, all experiments in this work are performed with fiducial pair reduction. To verify that this simplification does not introduce deviations, we also report an experiment without reduction here. We find similar results, regardless of whether we use fiducial pair reduction or not (column 4 vs. column 2). Note that running GST without fiducial pair reduction for two-qubit GST is too time-consuming.

Next, in column 5, we consider what happens when we operate the electron with simple $\pi/2$ pulses instead of the composite $\pi/2$ gate with XY8s for the nitrogen in a mixed state of all three levels (Fig. \ref{figs5}b, column 5). Compared to Fig. 3b of the main text (left-hand side) we find a significantly reduced fidelity for the $\pi/2$ pulse when the nitrogen is mixed compared to when it is initialised in $m_I = 0$. Compared to Fig. 3b of the main text (right-hand side) we see that applying XY8s around a $\pi/2$ pulse helps to improve the average gate fidelity in this case. 

One effect of the mixed nitrogen spin is that the electron spin picks up a different phase between pulses depending on the nitrogen spin state. To counteract this effect, we can either apply XY8 decoupling sequences or we can make sure the time between two consequent $\pi/2$ pulses is a multiple of the electron-nitrogen interaction time. The latter case is shown in the final column of Fig. \ref{figs5}b. We find no significant improvement compared to the results in column 5. One explanation could be that the native time between consequent $\pi/2$ pulses in column 5 is 1344 ns, which is already a close multiple of the electron-nitrogen interaction time.

\subsection{Single-qubit GST on the nitrogen spin}

\begin{figure*}[h]
\includegraphics[width=\textwidth]{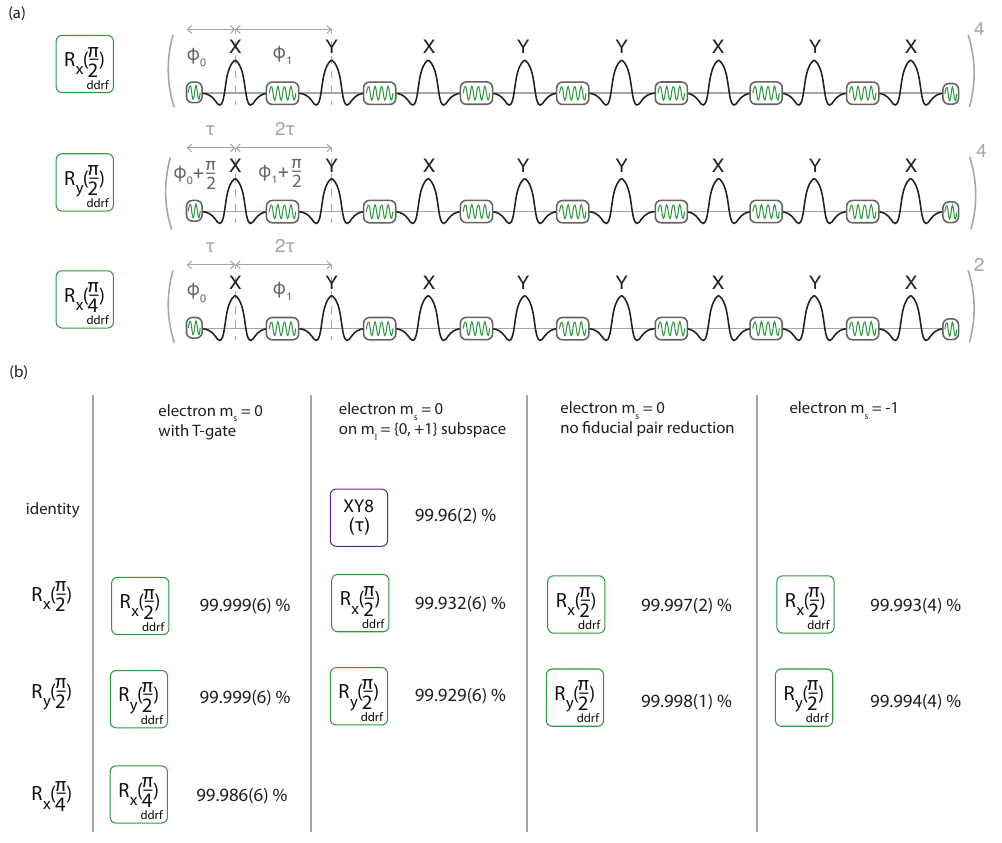}
\caption{\textbf{Additional nitrogen single-qubit GST results.} \textbf{a.} The collection of nitrogen gates applied in the various GST experiments. Next to the DDRF $\pi/2$ gates shown in the main text, we implement a T-gate ($\pi/4$ gate) by halving the number of DDRF units and without doing any further amplitude calibration. \textbf{b.} Average gate fidelities for the additional nitrogen single-qubit GST results. (column 1) Next to the previously characterised $\pi/2$ gates (Fig. 3), we characterise a DDRF $\pi/4$ gate. (column 2) Nitrogen gates on the $m_I = \{0, +1\}$ subspace, used in the nitrogen SWAP initialisation (Section \ref{nitrogen_init}). (column 3) Nitrogen gates without fiducial pair reduction. (column 4) Nitrogen gates with the electron in $m_s = -1$ instead of $m_s = 0$.}
\label{figs6}
\end{figure*}

Similarly to the electron spin, we implement a T-gate or $\pi/4$ gate on the nitrogen spin using the DDRF gate. We do not do any separate calibrations, but simply half the number of DDRF units for the $\pi/4$ gate compared to the $\pi/2$ gate (Fig. \ref{figs6}a). We find a comparable but somewhat worse fidelity for the $\pi/4$ gate, which could be due to the lack of separate amplitude calibration for that gate.

Next, we characterise the single-qubit gates on the $m_I = \{0, +1\}$ subspace of the nitrogen spin as opposed to the $m_I = \{0, -1\}$ subspace that we mostly discuss. Note that the $m_I = \{0, +1\}$ gates are used in the SWAP initialisation of the spin-1 nitrogen spin (Section \ref{nitrogen_init}). Instead of RF driving at the $m_I = 0, -1$ transition with the electron in $m_s = -1$, we now drive at the $m_I = 0, +1$ transition with the electron in $m_s = -1$, at a frequency of 2.780105 MHz. We find fidelities exceeding $99.9 \%$ for the nitrogen $\pi/2$ gates in the $m_I = \{0, +1\}$ subspace. Compared to the gates on the $m_I = \{0, -1\}$ subspace, these fidelities are significantly worse. The origin of this is in the value of $\tau = 7.304$ \textmugreek s, which has been optimised for the $m_I = \{0, -1\}$ subspace to avoid any unwanted rotations on the nitrogen spin during the application of an XY8 sequence on the electron spin (Section \ref{B_field}). For the gates on the $m_I = \{0, +1\}$ subspace, we also use $\tau = 7.304$ \textmugreek s, thereby introducing unwanted rotations on the nitrogen spin during the DDRF gate. We partly counteract this by calibrating the amplitudes for the $x$ and $y$ gates separately, but still find a worse fidelity than the gates on the $m_I = \{0, -1\}$ subspace. Further optimisation of the gates on the $m_I = \{0, +1\}$ subspace is therefore possible, for example by adjusting $\tau$ to a more suitable value. But, this is not pursued here.

Next, we characterise the single-qubit gates on the nitrogen spin without fiducial pair reduction (Fig. \ref{figs6}b, third column). We find comparable fidelities to the results obtained with fiducial pair reduction (Fig. 3). Finally, we consider the nitrogen DDRF gate with the electron in $m_s = -1$ instead of $m_s = 0$ (Fig. \ref{figs6}b, last column, see also Fig. 3). We find no discernible difference, suggesting that the nitrogen DDRF gate is a suitable gate for the regime where the electron spin is in an unknown or mixed state.

\clearpage 

\section{Additional two-qubit GST results}

\begin{figure*}[h]
\includegraphics[width=\textwidth]{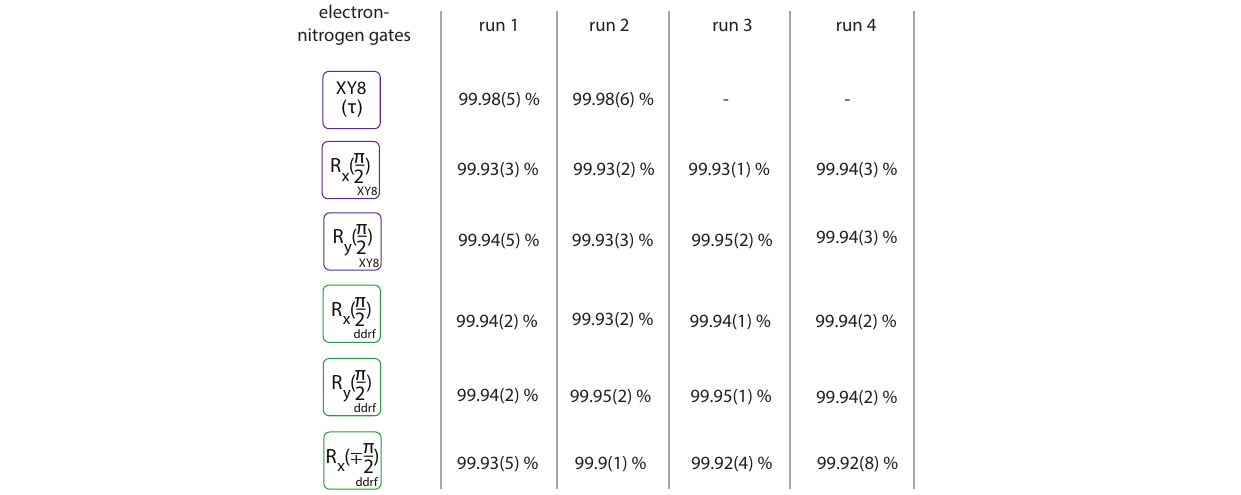}
\caption{\textbf{Average gate fidelities of all four two-qubit GST experiments.} We perform a total of four two-qubit GST experiments, two of which include the identity gate in the gate set (run 1, 2), and two of which do not include the identity gate (run 3, 4). All four runs were measured in a timespan of 9 days. Runs 1 and 2 take 16 hours each, and runs 3 and 4 take 14 hours each. We find consistent results across all four runs.}
\label{figs17}
\end{figure*}

In this section, we show the results of three additional two-qubit GST experiments, beyond the one presented in the main text. In Figure \ref{figs17} an overview is given. The results under run 1 are shown in the main text. The other three runs use the same gates, but for run 3 \& 4 the identity gate has been excluded from the gate set. We consistently measure average gate fidelities exceeding $99.9 \%$ for all gates, demonstrating the reproducibility of our results, as well as the robustness of our calibration routines and stability of our system considering the four runs were performed over a timespan of 9 days, with 16 hours each for run 1 \& 2, and 14 hours each for run 3 \& 4. 

We can calculate the weighted two-qubit gate fidelity of the four runs as follows:

\begin{equation}
    F_{avg} = \frac{\sum_i F_i / \sigma_i^2}{\sum_j 1 / \sigma_j^2},
\end{equation}

where $F_i$ is the fidelity for run $i$ and $\sigma_i$ is the standard deviation for run $i$. We obtain the error on the weighted average as:

\begin{equation}
    \sigma_{avg} = \sqrt{\frac{1}{\sum_i 1/\sigma_i^2}}.
\end{equation}

We obtain $F_{avg} = 99.923 \pm 0.026 \%$.

\newpage
\section{GST settings and model violation}

In this section, we show the settings for all experimental results with gate set tomograhpy presented in this work (Table \ref{gst_settings}). We group the results per figure and indicate the corresponding name in that figure. First, we show the maximum circuit depth $L$ used for that experiment. When we indicate $L = 128$, that implies that germs of depth $1, 2, 4, 8, 16, 32, 64$ and $128$ were run \cite{Nielsen2021}. Then, we indicate the experimental repetitions used for each circuit. The estimation error of gates in GST is typically of the order $O(1/L\sqrt{N})$ where $N$ is the number of repetitions \cite{Nielsen2021}. Lastly, we report the $N_\sigma$ metric for $L=128$, which quantifies the model violation \cite{Nielsen2021}. The error bar on the average gate fidelities obtained with GST represent one standard deviation (a $67 \%$ confidence interval). Another important element of the GST experiments is the preparation fiducials, germs and measurement fiducials used. We provide these separately, together with the data.

\begin{table}[h]
\begin{tabular}{|c|c|c|c|} 
 \hline
 \multicolumn{4}{|c|}{\textbf{Fig. S20}} \\ \hline 
name & L & repetitions & $N_\sigma$ \\ \hline 
 nitrogen $m_I = 0$ with T-gate & 128 & 1000 & 38.3 \\ \hline
nitrogen $m_I = 0$ & 128 & 1000 & -0.75 \\ \hline
nitrogen $m_I = 0$ $\tau = 2.0$ \textmugreek s & 128 & 1000 & 2.81 \\ \hline
nitrogen $m_I = 0$ no fiducial pair reduction & 128 & 1000 & 19 \\ \hline
nitrogen mixed no xy8 & 128 & 1000 & 49.8 \\ \hline
nitrogen mixed timing between pulses: 3208 ns & 128 & 1000 & 28.2 \\ \hline \hline
 \multicolumn{4}{|c|}{\textbf{Fig. S21}} \\ \hline 
 name & L & repetitions & $N_\sigma$ \\ \hline 
electron $m_s = 0$ with T-gate & 128 & 1000 & 4.9 \\ \hline
electron $m_s = 0$ on $m_I = \{0, +1\}$ subspace & 128 & 1000 & 1.79 \\ \hline
electron $m_s = 0$ no fiducial pair reduction & 128 & 1000 & -0.02 \\ \hline
electron $m_s = -1$ & 128 & 1000 & 0.72 \\ \hline \hline
 \multicolumn{4}{|c|}{\textbf{Fig. S22}} \\ \hline 
name & L & repetitions & $N_\sigma$ \\ \hline 
 run 1 & 128 & 1000 & 11.7 \\ \hline
 run 2 & 128 & 1000 & 15 \\ \hline
 run 3 & 128 & 1000 & 10.9 \\ \hline
 run 4 & 128 & 1000 & 15.4 \\ \hline
\end{tabular}
\caption{\textbf{Settings and model violation for all GST experiments.} Grouped per figure, we show the settings and model violation for every GST experiment in this work. $L$ is the maximum circuit depth, repetitions is the number of experimental repetitions per circuit and $N_\sigma$ quantifies the model violation \cite{Nielsen2021}.}
\label{gst_settings}
\end{table}

\clearpage
\section{Error decomposition: coherent vs. incoherent error}

Coherent and incoherent error processes do not contribute with equal weight to the metric of average gate fidelity. Coherent errors only contribute quadratically
to the average gate fidelity, whereas stochastic errors contribute linearly \cite{Nielsen2002,Sanders2016,Blume-Kohout2017}. Therefore, we argue that the measured average gate fidelities are mostly limited by single-qubit dephasing processes. In this section, we illustrate how the two different error types --- predicted by the GST error generators --- contribute to the measured average gate fidelity. 

In the GST framework \cite{Blume-Kohout2022}, we write the imperfect gate $G$ as the unitary target gate $G_0$ followed by a post-gate error process $\mathcal{E} = e^L$
\begin{align}
    G = e^L G_0,
\end{align}
where the post-gate error generator $L$ is a faithful representation of the magnitude and nature of errors in $G$. This error-generator $L$ is decomposed into elementary error generators with real coefficients
\begin{align}
    L = L_{H} + L_{S} + L_{C} + L_{A},
\end{align}
namely the Hamiltonian ($L_H$) and stochastic ($L_S$) error generators, as shown in Fig. 2 of the main text, the stochastic correlation ($L_C$) and the active or antisymmetric generator ($L_A$).

The average gate fidelity (see Methods) then is:
\begin{align}
   F_{\text{avg}} = \frac{\text{tr}(G^{\dagger}_{\text{exp}}G_{\text{0}})/d+1}{d+1}
    = \frac{\text{tr}(G^{\dagger}_{\text{0}}(e^L)^\dagger G_{\text{0}})/d+1}{d+1} .
\end{align}
In the regime of small errors ($e^L \rightarrow 1$), we can approximate this error process \cite{Blume-Kohout2022}: 
\begin{align}
    e^L = \sum_{k=0} \frac{L^k}{k!} = \mathds{1} + L + O(L^2) .
\end{align}
In this limit, the following equality holds:
\begin{align}
    1 - F_{\text{avg}} = \left(1-F_H\right) + \left(1-F_{\text{rest}}\right),
\end{align}
which we can use to distinguish between coherent errors caused by $L_H$ and the remaining incoherent types of errors ($L_{\text{rest}} = L_S+L_C+L_A$).

\begin{figure*}[h]
\includegraphics[width=0.7\textwidth]{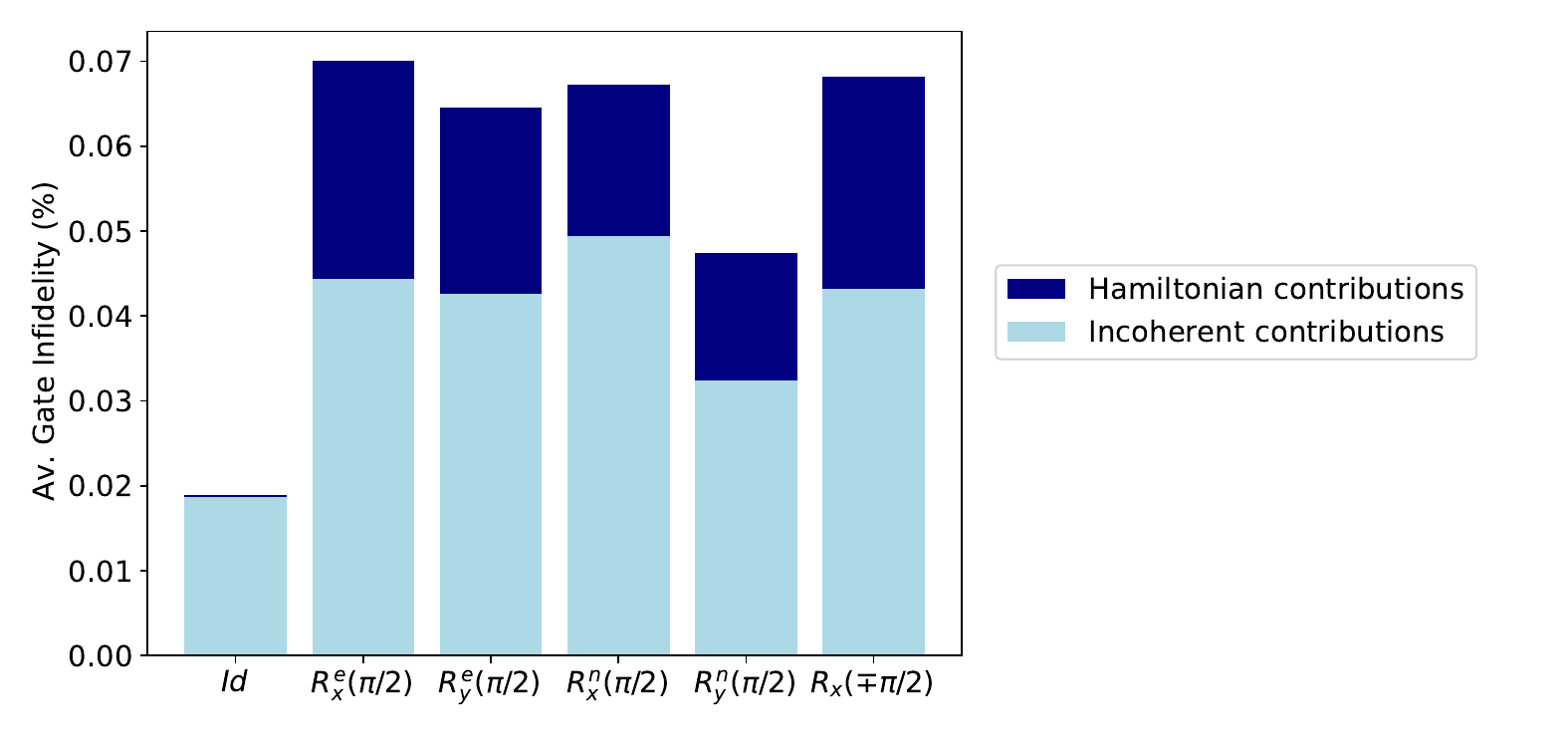}
\caption{\textbf{Coherent and incoherent contributions to the measured average gate infidelities.} Using the error generator from the GST experiment in Fig. 2 of the main text, we show how much coherent (Hamiltonian) and incoherent error processes contribute to the average gate infidelity. This underlines the argument that we are mostly limited by electron single-qubit dephasing.}
\label{fig:figs23}
\end{figure*}

In Fig. \ref{fig:figs23}, we compare the different contributions to the average gate fidelity. This underlines that in our case the Hamiltonian errors are not dominating the infidelities. In the case of the controlled two-qubit gate, the most prominent process is single-qubit dephasing of the electron spin (Fig. 2e). 

\clearpage
\bibliography{supp.bib}